\documentclass[journal ]{new-aiaa}
\usepackage[utf8]{inputenc}

\usepackage{amsmath, amsthm}
\usepackage{textcomp}

\usepackage{graphicx}
\usepackage[version=4]{mhchem}
\usepackage{siunitx}
\usepackage{longtable,tabularx}
\usepackage{subcaption}
\usepackage{float}
\usepackage[margin=1in]{geometry}
\usepackage{tikz}
\usepackage{comment}
\usepackage{cleveref}
\usepackage{graphicx}

\usetikzlibrary{shapes,arrows,positioning}

\title{Adjoint-Based Aerodynamic Shape Optimization \\ with a Manifold Constraint Learned by Diffusion Models}

\author{Long Chen\footnote{Postdoc (Habilitation candidate).}, Emre Özkaya\footnote{Senior researcher.}, Jan Rottmayer\footnote{Ph.D. student.}, and Nicolas R. Gauger\footnote{Professor, Associate Fellow AIAA.}}
\affil{Chair for Scientific Computing, University of Kaiserslautern-Landau (RPTU) \\ Bldg 34, Paul-Ehrlich-Strasse, 67663 Kaiserslautern}
\author{Zebang Shen\footnote{Lecturer.}}\affil{Institute for Machine Learning, ETH Zürich\\ Andreasstrasse 5, 8092 Zürich}
\author{Yinyu Ye\footnote{Professor Emeritus.}}
\affil{Department of Management Science and Engineering and ICME, Stanford University \\ Huang Engineering Center 308,  CA 94305-4121}

\newcommand{\cN}{\mathcal{N}}
\newcommand{\cM}{\mathcal{M}}
\newcommand{\cT}{\mathcal{T}}
\newcommand{\cP}{\mathcal{P}}
\newcommand{\rmd}{\mathrm{d}}
\newtheorem*{prediction*}{Prediction}

\begin{document}

\maketitle

\begin{abstract}

Significant advances in adjoint methods have enabled efficient and accurate computation of shape gradients for high-dimensional and high-fidelity aerodynamic shape optimization problems. However, the downstream optimization task, i.e., using the adjoint gradients to find optimal shape designs, still faces substantial challenges. The strong nonlinearity and nonconvexity of flow-physics-induced optimization landscapes complicate the search process of gradient-based algorithms, and trapping in local solutions is both theoretically unavoidable and difficult to control. Furthermore, as standard optimization solvers are formulated based purely on mathematical models, they cannot account for implicit constraints imposed by flow physics or practical computational considerations. These difficulties force practitioners to rely on trial-and-error tuning and variable scaling that undermines the efficiency promised by the adjoint method. To address these limitations, we introduce an adjoint-based optimization framework that integrates a diffusion model trained on existing designs to learn a smooth manifold of aerodynamically viable shapes. This manifold is enforced as an equality constraint to the shape optimization problem. Central to our method is the computation of adjoint gradients of the design objectives (e.g., drag and lift) with respect to the manifold space. These gradients are derived by first computing shape derivatives with respect to conventional shape design parameters (e.g., Hicks-Henne parameters) and then backpropagating them through the diffusion model to its latent space via automatic differentiation. Our framework preserves mathematical rigor and can be integrated into existing adjoint-based design workflows with minimal modification. Demonstrated on extensive transonic RANS airfoil design cases using off-the-shelf and general-purpose nonlinear optimizers, our approach eliminates ad hoc parameter tuning and variable scaling, maintains robustness across initialization and optimizer choices, and achieves superior aerodynamic performance compared to conventional approaches. This work establishes how AI generated priors integrates effectively with adjoint methods to enable robust, high-fidelity aerodynamic shape optimization through automatic differentiation.

\end{abstract}

\section{Introduction}

\lettrine{O}{ptimization} challenges in computational engineering and sciences are typically addressed through two complementary strategies. The first strategy designs tailored optimization algorithms for specific problem classes; prominent examples include the Simplex and Interior-Point Methods for linear programming, Optimality Criterion and Method of Moving Asymptotes for topology optimization, Stochastic Gradient Descent and its advanced variants for neural network training. The second strategy remodels the optimization problem formulation itself, i.e., mathematically or numerically reformulating it to achieve comparable optima with reduced computational complexity to solve. This includes techniques like reduced-order modeling, surrogate modeling, problem-specific parameterization (e.g., Hicks-Henne airfoil parameterization), convex relaxation (e.g., Semidefinite Programming for NP-hard problems). For adjoint-based aerodynamic shape optimization, the first strategy faces inherent difficulties stemming from high nonlinearity and nonconvexity of the optimization problem (due to highly complex flow physics), coupled with computationally expensive black-box function and gradient evaluations. Devising tailored optimization algorithms is challenging both theoretically and empirically. Consequently, researchers and practitioners predominantly rely on off-the-shelf and general-purpose nonlinear optimization solvers such as SLSQP, IPOPT, and SNOPT. This limitation has redirected significant research effort toward the remodeling strategy, particularly through advanced shape parameterization and model-order reduction techniques. Leveraging recent advances of powerful generative AI models, we introduce a novel remodeling approach for adjoint-based aerodynamic shape optimization: restricting the design space of the optimization problem via a manifold constraint learned by a diffusion model.

\subsection{Aerodynamic Shape Optimization Problem Formulation with Manifold Constraint}

\color{black}
Aerodynamic shape optimization (ASO) has emerged as a cornerstone of modern aircraft design, enabling systematic improvements in performance metrics such as lift-to-drag ratio, stability, and fuel efficiency. Substantial research efforts have been put into parameterization of the aerodynamic shape, which serves as the foundation for exploring the design space. An effective parameterization balances geometric flexibility with numerical stability, thereby facilitating the identification of high-performing configurations within practical computational limits.
However, a fundamental conflict arises in the optimization process. On one hand, high-fidelity parameterizations with a large number of design variables are desirable to capture subtle geometric variations and expand the solution space, offering the potential to discover superior aerodynamic shapes. On the other hand, increasing the dimensionality of the design space introduces significant optimization challenges. The optimization landscape becomes more complex, often resulting in ill-conditioning and slower convergence. These competing demands necessitate careful methodological choices in both parameterization and optimization strategy to ensure computational efficiency and robustness. This challenge is particularly acute in adjoint-based optimization, which targets high-dimensional problems.

The starting point is the generic aerodynamic shape optimization problem:
\begin{align}\label{eq:aso}
    \text{(ASO)}\quad \left\{
    \begin{array}{ll}
        \min\limits_{x} & J(u(x), x) \\
        \text{subject to} & R(u(x), x) = 0 \quad \text{(state equation)} \\
                          & c_i(u(x), x) \leq 0, \quad i = 1, \ldots, m
    \end{array}
    \right.
\end{align}
Here, \( J: \mathbb{R}^n \times \mathbb{R}^d \rightarrow \mathbb{R} \) is the objective function, and \( c: \mathbb{R}^n \times \mathbb{R}^d \rightarrow \mathbb{R}^m \) denotes the inequality constraints. The vector \( x \in \mathbb{R}^d \) represents the design parameters in a \( d \)-dimensional Euclidean space, while \( u \in \mathbb{R}^n \) corresponds to the state variables. The state equation \( R(u(x), x) = 0 \) typically represents a system of partial differential equations (PDEs), ensuring that the state vector \( u \) corresponds to a physically valid flow solution. In aerodynamic shape optimization, this PDE is often the Navier–Stokes or Reynolds-Averaged Navier–Stokes (RANS) equations.

Note that even for moderate values of \( d \), the design space \( \mathbb{R}^d \) is vast and may encompass many unrealistic or non-physical geometries. Exploring such a high-dimensional space introduces significant challenges in practical shape optimization. In particular, the optimization landscape tends to become increasingly ill-conditioned due to intricate parameter interactions and nonlinear dependencies. As the number of design variables increases, the objective function may develop narrow, curved valleys, flat regions, or steep ridges, which complicate the convergence of gradient-based methods. This ill-conditioning is typically reflected in the Hessian matrix through a large condition number, leading to descent directions that are poorly aligned with the true optimal path. This problematic is widely observed across different aerodynamic shape optimization problems and advanced geometric parameterization methods were developed \cite{koo2018investigation} \cite{streuber2021dynamic} \cite{wu2022sensitivity} \cite{kedward2020gradient} \cite{bons2020aerostructural}. Moreover, being designed purely mathematically, general-purpose optimization algorithms are unable to account for implicit constraints arising from flow-physics and computational practice. An optimizer may propose shape updates that are mathematically perfectly valid but lead to less accurate or even non-convergent flow solutions due to their irregular shapes. As a result, ad hoc tuning of optimization problem parameters is often necessary for adjoint-based ASO. For a concrete illustration, see the different scalings for objective and constraint function and gradient in the SU2 design optimization tutorial test cases \cite{su2_tutorials}.

Over the past years, several efforts have been made to address the challenge of high-dimensional design spaces through dimensionality reduction techniques. One notable approach is the use of Proper Orthogonal Decomposition (POD), which extracts dominant shape modes from a database of geometries; this technique, when combined with an interpolation model, has been effectively applied by Iuliano et al. \cite{IULIANO2013327} to aerodynamic shape optimization. In another line of research, Bouhlel et al. \cite{BouhlelKPLS} employed Partial Least Squares (PLS) regression to identify the most influential directions in the design space, enabling a reduced set of parameters to be used in a Bayesian optimization framework. These methods aim to retain the most relevant design variations while alleviating the numerical difficulties associated with large parameter sets.

While these dimensionality reduction methods have demonstrated promising results, they also come with inherent limitations. Techniques such as POD and PLS typically rely on linear assumptions and may fail to fully capture the complex, non-linear relationships between design parameters and aerodynamic performance metrics. As a result, important non-linear interactions may be overlooked, potentially restricting the expressiveness and accuracy of the reduced design space. This limitation becomes especially critical in high-fidelity aerodynamic optimization problems, where non-linear effects often dominate the system's behavior. Consequently, there remains a need for more advanced reduction techniques that can preserve non-linearity while maintaining computational efficiency.

A promising strategy to address these challenges is to restrict the optimization process to a lower-dimensional manifold embedded within the original design parameter space. This manifold represents a more regular and physically meaningful subset of \( \mathbb{R}^d \), thereby eliminating unrealistic geometries from the search space. By constraining the design parameters to vary within this manifold, the optimization landscape becomes smoother and more well-behaved, allowing gradient-based methods to operate more effectively and efficiently. We formulate a manifold-constrained aerodynamic shape optimization (MASO) problem as

\begin{align}\label{eq:mcaso}
    \text{(MASO)}\quad \left\{
    \begin{array}{ll}
        \min\limits_{x} & J(u(x), x) \\
        \text{subject to} & R(u(x), x) = 0 \quad \text{(state equation)} \\
                          & c_i(u(x), x) \leq 0, \quad i = 1, \ldots, m \\
                          & x \in \mathcal{X},
    \end{array}
    \right.
\end{align}
where $\mathcal{X}$ is the manifold that is a subset of $\mathbb{R}^d$, and $x \in \mathcal{X}$ is the only extra constraint added to the orignal ASO problem \eqref{eq:aso}. This reformulation has the advantage that it does not change the parameterization $x$ and is therefore flexible and easy to integrate into existing shape optimization framework. The question now to be addressed is how to design and construct such a manifold $\mathcal{X}$ that optimally constrains our design space.

\subsection{Learning Manifold Constraint with Diffusion Models}

The usefulness of the reformulation \eqref{eq:mcaso} to the adjoint-based aerodynamic shape optimization problem highly depends on the manifold constraint, which should satisfy three requirements: First, the manifold must be \textit{differentiable}. Second, the manifold must be \textit{low-rank}, i.e., restricting designs exclusively to physically and aerodynamically functional geometries. Third, it must ensure \textit{richness} by containing optimal solutions across diverse aerodynamic scearios (e.g., varying Mach numbers, Reynolds numbers) and different design objectives and constraints. However, the last two requirements lack precise mathematical formalization due to inherent ambiguities: Physical and aerodynamical viable shapes lacks universal analytical criteria; Optimality can only be validated within specific design problem contexts. Hence, no explicit and first-principle based formulation of such a manifold constraint is possible. In this work, we propose  a data-driven approach that leverages the recent advancement in generative diffusion model for the manifold construction. 

Generative diffusion models resolve this challenge by learning an implicit representation of such a manifold from existing high-performing designs. In this work, we establish the link between a desirable manifold constraint $x \in \mathcal{X}$ in \eqref{eq:mcaso} and diffusion models through the following three points:
\begin{itemize}
    \item [1.] By training on existing aerodynamic shapes, the diffusion model implicitly learns the underlying data distribution $p_{data}$ of aerodynamically viable shapes. The generative process is a sampling process that generates/samples a design $x \in \mathbb{R}^d$ sampled from this learned data distribution,
    \begin{equation}
        x = G_\theta(z), 
    \end{equation}
    where $z \in \mathbb{R}^d$ is the so-called latent variable and is sampled from a Gaussian distribution, i.e., $z \sim \mathcal{N}(\mathbf{0},\mathbf{I})$, and $\theta$ is the trainable parameter of the diffusion model. Let $\mathcal{X}_G$ denotes the set of shape parameters $x$ generated by $G_\theta$,
    \begin{equation}
    \mathcal{X}_G := \{ x: x = G_\theta(z), z \in \mathcal{Z} \},
    \end{equation}
    where $\mathcal{Z}$ is some bounded set centered around the mean of the input Gaussian distribution. Intuitively, the generative process $G_\theta$ of the diffusion model maps the latent space $\mathcal{Z}$ to the learned data manifold $\mathcal{X}_G$. If the training data is aerodynamically viable, then the learned data manifold is expected to be \textit{low-rank} that excludes unrealistic geometries (see Section \ref{subsec:diffusion_model} and \ref{sec:learned_manifold}).
    \item [2.] We derive and implement a \textit{differentiable} framework for computing the adjoints of the manifold-constrained aerodynamic shape optimization problem \eqref{eq:mcaso}. Central to our method is the computation of adjoint gradients of the design objectives with respect to the manifold space. These gradients are derived by first computing shape derivatives with respect to conventional shape design parameters and then backpropagating them through the diffusion model to its latent space via Automatic Differentiation (see Section \ref{subsec:adjoint_so} and \ref{subsec:adjoint_diffusion_manifold}). 
    \item [3.] As a state-of-the-art generative AI model, diffusion model demonstrates excellent ability to generalize on complex high-dimensional data distributions \cite{ho2020denoising}\cite{stanczuk2024diffusion}. In this work, we performed extensive systematic experiments to empirically demonstrate that, when trained on a sufficiently diverse set of aerodynamic shapes, the learned manifold contains optimal solutions $x^\star$ with very high probability, and hence ensures \textit{ richness} (see Section \ref{sec:results}). 
\end{itemize}

\begin{figure}
    \centering
    \input{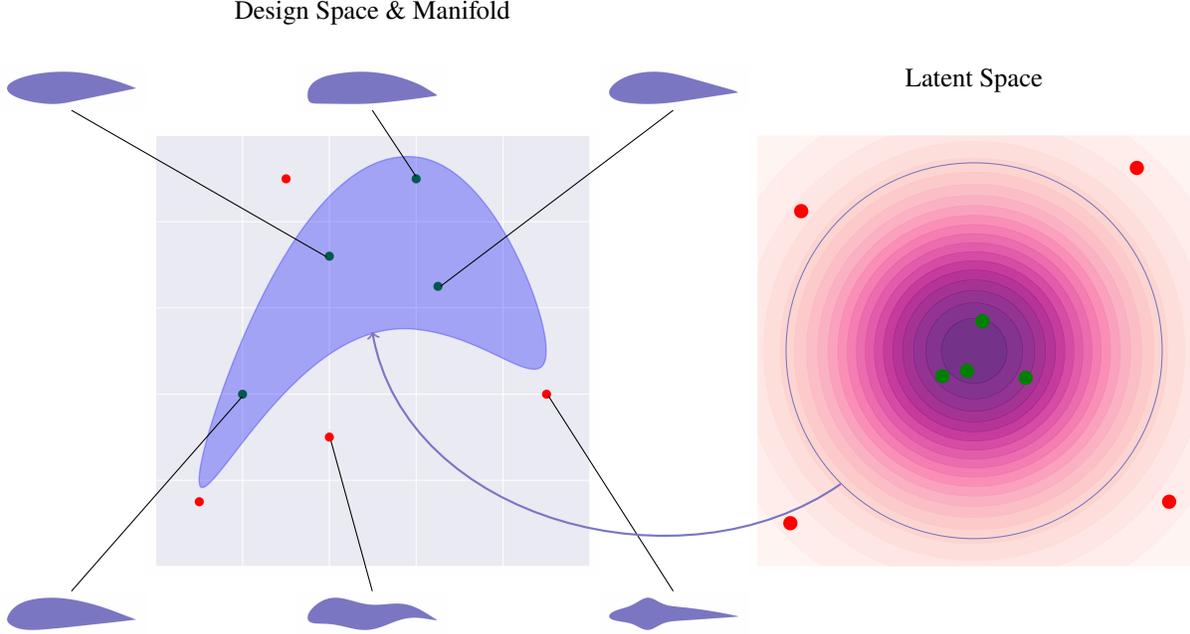}
    \caption{Illustration of a parameter manifold $\mathcal{X}_G$ embedded in the parameter space X. Arrows denote the shape decoding processes.}
    \label{fig:diffusionairfoil_manifold}
\end{figure}

Computational experiments show that our proposed approach eliminates the need for ad hoc parameter tuning and variable scaling across different aerodynamic shape optimization problems and optimization solvers. The optimization process is robust to the choice of initialization, and the resulting airfoil designs consistently achieve superior aerodynamic performance compared to their conventional counterpart, often converging more quickly. These promising results empirically validate the effectiveness of our proposed manifold-constrained ASO framework, and highlight how AI-generated design priors can synergize with adjoint-based formulations to enable robust, high-fidelity aerodynamic shape optimization.

\subsection{Related Works on Using Diffusion Models for Aerodynamic Shape Design}

Recently, several studies have applied diffusion models to aerodynamic shape design in various aspect. Wei et al. \cite{wei2024diffairfoil} proposed an airfoil sampling method based on a latent space diffusion model combined with an auto-decoder model, demonstrating its advantages over Generative Adversarial Networks (GANs) through extensive comparisons. Graves et al. \cite{graves2024airfoil} introduced a conditional diffusion model for airfoil shape generation that directly uses surface coordinates as training data. Wagenaar et al. \cite{wagenaar2024generative} trained conditional diffusion models for airfoil generation and showed that they produce diverse candidate designs under identical requirements and constraints, effectively exploring the design space to provide multiple starting points for optimization procedures. Our findings regarding the superiority of diffusion models over other generative models are consistent with the literature above. For instance, using the same training data, we were unable to learn a comparable manifold with GANs. Unlike these earlier works, our study focuses on adjoint-based shape optimization. To the best of our knowledge, this is the first work to propose using diffusion models to learn a differentiable manifold constraint in aerodynamic shape optimization. This manifold constraint can easily be integrated into existing adjoint-based shape design frameworks. By leveraging the powerful generative capabilities of diffusion models, the developed differentiable framework substantially enhances adjoint-based aerodynamic shape optimization.

\section{Method}
\label{sec:method}

\subsection{Adjoint-Based Shape Optimization}
\label{subsec:adjoint_so}

In aerodynamic design optimization, the choice of optimization strategy is closely tied to the number of design variables. For problems with a moderate number of parameters, gradient-free methods such as Efficient Global Optimization (EGO) or evolutionary algorithms are often preferred due to their robustness and global search capabilities. However, these methods suffer from poor scalability and become computationally infeasible as the number of design parameters increases. In high-dimensional settings, gradient-based optimization becomes more suitable thanks to its superior scalability and efficiency. The challenge then shifts to the computation of the gradient vector, which can be extremely costly if naively approached—especially when the number of design variables is large. This is where the adjoint methodology becomes essential. The adjoint approach enables efficient computation of sensitivities with a computational cost that is largely independent of the number of parameters, making it particularly attractive for large-scale optimization problems.

There are two principal forms of the adjoint method: the continuous and the discrete approach. The continuous adjoint is derived from the governing equations before discretization and can offer analytical clarity and computational benefits, especially for structured grids. However, it may suffer from inconsistencies and become difficult to maintain within complex numerical solvers. The discrete adjoint \cite{GilesPierce2000}, on the other hand, is constructed directly from the discretized equations used in the solver, ensuring exact consistency with the numerical output. Furthermore, the use of Algorithmic Differentiation (AD) facilitates the development of discrete adjoint solvers with minimal manual intervention and high computational efficiency, making them particularly well-suited for modern aerodynamic design workflows.
\subsubsection{Discrete Adjoint}

We reconsider the generic shape optimization problem \eqref{eq:aso}. To solve it using a typical gradient-based optimization method, we require the gradient of the objective function with respect to the design variables. The total derivative of \( J \) with respect to \( x \) is given by:

\begin{equation}
    \frac{dJ}{dx} = \frac{\partial J}{\partial x} + \frac{\partial J}{\partial u} \frac{du}{dx} .
    \label{eq:dJ_dx}
\end{equation}

Here, the total derivative \( \frac{dJ}{dx} \in \mathbb{R}^d \), and the partial derivatives \( \frac{\partial J}{\partial x} \in \mathbb{R}^d \), \( \frac{\partial J}{\partial u} \in \mathbb{R}^n \) are row vectors. The term \( \frac{du}{dx} \) is the Jacobian matrix of size \( n \times d \), which is generally infeasible to evaluate directly.

We can reformulate the state constraint as a fixed-point equation:

\begin{equation}
    R(u(x), x) \implies u = F(u, x),
\end{equation}

where \( F \) is a differentiable fixed-point operator that encapsulates all operations performed within a pseudo-time step of the primal PDE solver. Typically, one pseudo-time step involves spatial discretization schemes and preconditioning procedures.
 From the fixed-point iteration we get:
\begin{equation}
    \frac{du}{dx} = \frac{\partial F}{\partial u} \frac{du}{dx}  + \frac{\partial F}{\partial x}.
\end{equation}
From the above equation, we get
\begin{equation}
 \frac{du}{dx}  = \left( I - \frac{\partial F}{\partial u} \right)^{-1} \frac{\partial F}{\partial x}.
 \label{eq:du_dx}
\end{equation}
Using this in eq. \ref{eq:dJ_dx}, we get
\begin{equation}
    \frac{dJ}{dx} = \frac{\partial J}{\partial x} + \frac{\partial J}{\partial u}  \left( I - \frac{\partial F}{\partial u} \right)^{-1} \frac{\partial F}{\partial x} .
    \label{eq:dJ_dx2}
\end{equation}
Now we define the adjoint vector (column vector) $\lambda \in \mathbb{R}^n$ as
\begin{equation}
\lambda^T =  \frac{\partial J}{\partial u}  \left(  I - \frac{\partial F}{\partial u}  \right)^{-1},
\label{eq:adj_vector}
\end{equation}
or 
\begin{equation}
\lambda =  \left(  I - \frac{\partial F}{\partial u}  \right)^{-T}\frac{\partial J}{\partial u}^T  ,
\label{eq:adj_vector}
\end{equation}
From the above equation, we derive the fixed-point equation for the adjoint vector
\begin{equation}
 \lambda =  \frac{\partial J}{\partial u}^T  + \frac{\partial F}{\partial u}^T \lambda,
\label{eq:adj_fixed_point},
\end{equation}
which can be found iteratively by the adjoint fixed-point iterations:
\begin{equation}
 \lambda_{k+1} =  \frac{\partial J}{\partial u}^T  + \frac{\partial F}{\partial u}^T \lambda_k, \; k=0,\ldots
\label{eq:adj_fixed_point_scheme},
\end{equation}
Once the adjoint fixed-point iterations converge, the gradient vector can be evaluated by
\begin{equation}
    \frac{dJ}{dx} = \frac{\partial J}{\partial x} + \lambda_\ast^T \frac{\partial F}{\partial x} .
\end{equation}
or 
\begin{equation}
    \frac{dJ}{dx}^T = \frac{\partial J}{\partial x}^T +  \frac{\partial F}{\partial x}^T \lambda_\ast.
    \label{eq:dJ_dx_final_T}
\end{equation}

From an implementation standpoint, the most effective approach is to evaluate all expressions involving transposed Jacobian--column vector products using the reverse mode of Algorithmic Differentiation (AD). Recall that for any function \( y = f(x) \), the reverse mode computes \( \bar{x} = \left( \frac{\partial y}{\partial x} \right)^T \bar{y} \), where \( \bar{y} \) and \( \bar{x} \) are column vectors. 

Therefore, in practice, it suffices to apply AD techniques to differentiate the primal fixed-point scheme \( F \) and the post-processing function \( J \). Once this step is completed, all the expressions in Eqs.~\eqref{eq:adj_fixed_point_scheme} and \eqref{eq:dJ_dx_final_T} can be evaluated automatically using the backward sweep procedure of AD.

For the numerical experiments presented in this work, we employ the discrete adjoint capabilities of the open-source CFD framework \texttt{SU2} \cite{SU2_disc_adj}. The adjoint solver in \texttt{SU2} has been developed based on the methodology outlined above and leverages Algorithmic Differentiation (AD) through the use of the \texttt{CoDiPack} AD library \cite{SaAlGauTOMS2019}. Advanced techniques such as preaccumulation are employed to reduce both memory overhead and computational cost. The resulting discrete adjoint implementation in \texttt{SU2} is robust and efficient, supporting adjoint simulations of the Reynolds-Averaged Navier–Stokes (RANS) equations without resorting to common simplifications such as the frozen eddy viscosity assumption \cite{PETER2010373}, which is often used in other solvers. As a result, turbulence models are treated consistently within the adjoint framework. This enhances both the accuracy and reliability of computed sensitivities.

\subsubsection{Airfoil Shape Parameterization}

Our method is generally applicable to all shape parameterization techniques, as it does not rely on any specific representation of the geometry. The choice of shape parameterization is typically application-specific and is determined by the user prior to initiating an optimization study. In the present work, we adopt the well-established Hicks-Henne bump functions to deform the airfoil shape. This choice is in line with standard practice in the literature and is consistent with benchmark cases available in the SU2 test case repository. Various formulations of the Hicks-Henne method exist; here, we follow the definition used in SU2~\cite{economon2014optimal}, where the bump function is given by  
\begin{equation}
f_n(x) = \sin^3(\pi x^{e_n}), \qquad e_n = \frac{\log(0.5)}{\log(x_n)}, \qquad x \in [0, 1].
\end{equation}
This formulation ensures that the bump function attains its maximum at \( x_n \) and vanishes at the endpoints of the interval.
The total deformation of the airfoil surface from its original shape at an \( x \)-location along the chord is computed as 
\begin{equation}
\Delta y(x) = \sum_{n = 1}^N \delta_n f_n(x),
\end{equation}
where \( N \) is the number of bump functions used and \( \delta_n \) denotes the coefficient associated with the \( n \)-th bump function. If we denote the original shape by \( s_0(x) \), then the Hicks-Henne shape representation reads
\begin{equation}
s(\delta; x) = s_0(x) + \sum_{n = 1}^N \delta_n f_n(x).
\end{equation}
The \( x \)-locations for the bump functions are chosen equidistantly along the chord line. These deformations are applied to both the suction and pressure sides of the airfoil. The coefficients \( \delta_n \) serve as the optimization parameters in the present study.

\color{black}

\subsection{Diffusion Model}
\label{subsec:diffusion_model}

Diffusion model (DM) is a class of latent variable models that learns to generate data by reversing the diffusion process.
The main concept is to add noise to the data step by step, then train the model to reverse this process, enabling it to generate data from random noise during inference.
DM has achieved great success in various applications \cite{yang2023diffusion}.

Mathematically, DM approximates a given data distribution $q(x_0)$ with a parameterized family $p_\theta(x_0)$. 
Derived within the framework of the variational Bayesian method, a DM should specify two design choices: (1) A latent variable $z$ and the corresponding variational distribution (sometimes called the inference distribution) $q(z | x)$ and (2) a joint generative distribution $p_\theta(x, z)$.
With these two choices, DM can be trained by maximizing the standard evidence lower bound (ELBO).

\subsubsection{Denoising Diffusion Probabilistic Model}

In this work, we focus on the denoising diffusion probabilistic model (DDPM) \cite{ho2020denoising} and follow the choice of $q(z|x)$ and $p_\theta(x, z)$ therein.
Consider a forward diffusion process which is a Markov chain that gradually adds Gaussian noise to the data according to a variance schedule $\beta_1, \dots, \beta_T$:
\begin{equation}
    q(x_{1:T} | x_0) := \prod_{t=1}^T q(x_t | x_{t-1}),~~ q(x_t | x_{t-1}) := \mathcal{N}(x_t; \sqrt{1-\beta_t} x_{t-1}, \beta_t \mathbf{I}).
\end{equation}
DDPM sets the latent variable as $z = x_{1:T}$.
Note that the joint variational distribution of the trajectory $x_{1:T}$ conditioned on $x_0$ remains Gaussian.
This leads to two crucial observations when conditioned on $x_0$: Denoting $\alpha_t = 1-\beta_t$ and $\bar{\alpha}_t := \prod_{s = 1}^t \alpha_s$, we have
\begin{itemize}
    \item The forward process posterior $q(x_{t-1} | x_t, x_0)$ in DDPM is within the same Gaussian family, i.e.
    \begin{equation} \label{eqn_forward_posterior}
        q(x_{t-1}|x_{t}, x_0) = \mathcal{N}\left(x_{t-1}; \frac{\sqrt{\bar \alpha_{t-1}}\beta_t}{1 - \bar \alpha_t} x_0 + \frac{\sqrt{\alpha_t}(1-\bar \alpha_{t-1})}{1-\bar\alpha_t}x_t, \frac{(1-\bar \alpha_{t-1})}{1-\bar\alpha_t} \beta_t \mathbf{I}\right).
    \end{equation}
    \item Any marginal distribution $q(x_t | x_0)$ of the trajectory distribution $q(x_{1:T} | x_0)$ remains Gaussian. 
    \begin{equation} \label{eqn_marginal}
        q(x_t | x_0) = \mathcal{N} (x_t; \sqrt{\bar{\alpha}_t} x_0, (1-\bar{\alpha}_t) \mathbf{I}).
    \end{equation}
\end{itemize}
The joint generative distribution $p_\theta(x, z)$ of DDPM is defined as a \emph{Markov chain} with parameterized Gaussian transitions starting at $p(x_T) = \mathcal{N}(x_T; \mathbf 0, \mathbf I)$:
\begin{equation} \label{eqn_generative_transition}
    p_\theta(x_{0:T}) := p(x_T) \prod_{t=1}^T p^t_\theta(x_{t-1} | x_t) \text{ with } p^t_\theta(x_{t-1}|x_{t}) := \mathcal{N}(x_{t-1}; \mu_\theta(x_t, t), \frac{(1-\bar \alpha_{t-1})}{1-\bar\alpha_t}  \beta_t \mathbf{I} ).
\end{equation}
where the Gaussianity of the above transition is justified by the Gaussian form of the forward posterior (\ref{eqn_forward_posterior}).

Exploiting the Gaussian form of the marginal distribution (\ref{eqn_marginal}), with a simple linear reparameterization
\begin{equation}
    \mu_\theta(x_t, t) = \frac{1}{\sqrt{\alpha_t}}\left(x_t - \frac{\beta}{\sqrt{1-\bar \alpha_t}}\epsilon_\theta(x_t, t) \right),
\end{equation}
the training objective derived from ELBO can be equivalently described as predicting the noise added,
\begin{equation} \label{eqn_obj_DDPM}
L(\theta) := \mathbb{E}_{t, x_0, \epsilon}\left[\| \epsilon - \epsilon_\theta (\sqrt{\bar{\alpha}_t} x_0 + \sqrt{1 - \bar{\alpha}_t} \epsilon, t) \|^2 \right],
\end{equation}
where $\mathbb{E}$ is the expectation, $t$ is uniform between 1 and $T$, $\epsilon \sim \mathcal{N}(0, \mathbf{I})$, $x_0$ is sampled according to the empirical distribution of the dataset.

Note that for a given parameter $\theta$, DDPM can generate data from $p_\theta(x_0)$ by starting from $x_T$ sampled from $\mathcal{N}(x_T; \mathbf 0, \mathbf I)$ and follow the above transition (\ref{eqn_generative_transition}) for $t = T$ to $t=1$.

\subsubsection{Denoising Diffusion Implicit Model}

DDPM is a stochastic sampling process, i.e., $x= G_\theta(z)$ will yield a different $x$ each time, even when the same $z$ is used. A drawback of the DDPM's generation process is that it takes many steps to produce a high quality sample from the learned distribution.
To obtain a deterministic and fast generative process, we use the Denoising Diffusion Implicit Model (DDIM) sampling \cite{song2020denoising}. 
Similarly to DDPM, DDIM sets the latent variable $z = x_{1:T}$, but with the distinction that it constructs the variational distribution $q_\sigma(x_{1:T}|x_0)$ in a \emph{non-Markovian} manner
\begin{equation}
    q_\sigma(x_{1:T}|x_0) := \prod_{t=1}^T q_\sigma(x_{t}|x_{t-1}, x_0) = q_\sigma(x_T | x_0)  \prod_{t=2}^T q_\sigma(x_{t-1} | x_{t}, x_0)
\end{equation}
such that the corresponding marginal distribution $q_\sigma(x_t|x_0)$ exactly matches \cref{eqn_marginal} and the 
forward process posterior $q_\sigma(x_{t-1} | x_{t}, x_0)$ is of the form
\begin{equation}
    q_\sigma(x_{t-1} | x_{t}, x_0) = \mathcal{N}\left(\sqrt{\bar \alpha_{t-1}} x_0 + \sqrt{1 - \bar \alpha_{t-1} - \sigma_t^2}\cdot \frac{x_t - \sqrt{\bar \alpha_t}x_0}{\sqrt{1-\bar \alpha_t}}, \sigma_t^2 \mathbf{I}\right).
\end{equation}
Here, $\sigma\in\mathbb{R}^T$ is some noise scheduling and $q_\sigma(x_T | x_0) = \mathcal{N}(\sqrt{\bar \alpha_T}, (1-\bar \alpha_T) \mathbf{I})$.
The joint generative distribution $p_\theta(x, z)$ of DDIM is constructed to model the above forward process posterior with a parameterized ``predicted $x_0$''
\begin{equation}
   p_\theta(x_{0:T}) := p(x_T) \prod_{t=1}^T p^t_\theta(x_{t-1} | x_t) \text{ with } p_\theta^t(x_{t-1}|x_{t}) = q_\sigma(x_{t-1} | x_t, f_\theta(x_t, t)).
\end{equation}
Moreover, the authors of DDIM show that with the parameterization
\begin{equation}
    f_\theta(x, t) = \frac{x_t - \sqrt{1 - \bar \alpha_t}\epsilon_\theta(x, t)}{\sqrt{\bar \alpha_t}},
\end{equation}
the training objective of DDIM, derived from ELBO, matches exactly that of DDPM (\ref{eqn_obj_DDPM}).

In this work, we follow the common practice and set $\sigma_t = 0$ for all $t$, so that the sampling process of DDIM becomes deterministic, i.e., the mapping $x = G_\theta(z)$ is unique.
The update rule of $x_t$ can be simplified to 
\begin{equation}
    x_{t-1} = \mathcal{D}_t(x_t) := \sqrt{\bar \alpha_{t-1}}f_\theta(x_t, t) + \sqrt{1-\bar \alpha_{t-1}}\epsilon_\theta(x_t,t)
\end{equation}
As the composition of $\mathcal{D}_t$, the sampling process $G$ is defined as
\begin{equation}
\begin{aligned}
x_{t-1} &= \mathcal{D}_t(x_t)\quad (t=T,T-1,\dots,1),\\
G(x_T) &= \bigl(\mathcal{D}_1 \circ \cdots \circ \mathcal{D}_T\bigr)(x_T).
\end{aligned}
\label{eq:ddim_sampling}
\end{equation}

\subsection{Adjoint-Based Shape Derivatives on Diffusion Model Manifold}
\label{subsec:adjoint_diffusion_manifold}

\subsubsection{Computation of Adjoints via Reverse-Mode AD}

Formally, we denote the generative process as the mapping
\begin{equation}
    G_\theta: z \rightarrow x(z),
\end{equation}
which maps each latent space variable $z$ to a Hicks-Henne parameter. We can reformulate the manifold constrained parametric shape optimization problem \eqref{eq:mcaso} as 
\begin{align}\label{eq:mcaso2}
    \text{(MASO)}\quad\left\{ \begin{array}{lrclr}
     \min &  J(u(x(z)),x(z),z) &  \\
    \text{~s.t.} &  R(u(x(z)),x(z),z) & = & 0 &\\
    & c_i(u(x(z)),x(z),z) & \leq & 0 
    \end{array}\right.
\end{align}
The adjoint of the objective function $J$ with respect to the latent variable $z$ reads
\begin{equation}
    \frac{d J}{d z} = \underbrace{\frac{\partial J}{\partial x}}_{\text{SU2 adjoint}}\cdot\underbrace{\frac{\partial  x(z)}{\partial  z}}_{\text{Diffusion backprop}} + \underbrace{\frac{\partial J}{\partial z}}_{=0}.
\end{equation}
The second term on the right-hand side $\frac{\partial J}{\partial z}$ vanishes since there is no explicit dependency of the objective function J on the latent variable z. Hence,
\begin{equation}
    \frac{d J}{d z} = \frac{\partial J}{\partial x} \cdot \frac{\partial  x(z)}{\partial  z},
\label{eq:adjoint_dm}
\end{equation}
which corresponds to reverse-mode AD formulation,
\begin{equation}
    \frac{d J}{d z} \bar J = \frac{\partial J}{\partial x} \cdot \frac{\partial  x}{\partial  z} \bar J.
\label{eq:adjoint_dm2}
\end{equation}
Equivalently,
\begin{equation}
    \frac{d J}{d z}^T \bar J =   \frac{\partial  x}{\partial  z}^T \frac{\partial J}{\partial x}^T \bar J.
\label{eq:adjoint_dm3}
\end{equation}
Since we have a scalar output we have $\bar J = 1$, therefore we have
\begin{equation}
    \frac{d J}{d z}^T  =   \frac{\partial  x}{\partial  z}^T \frac{\partial J}{\partial x}^T,
\label{eq:adjoint_dm4}
\end{equation}
which is a vector Jacobian product that represents the \textit{canonical backward step} in reverse-mode AD. 

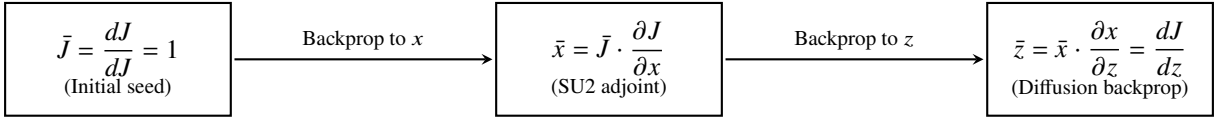
\begin{figure}[ht]
\centering
\begin{tikzpicture}[
    node distance=3.5cm,  
    block/.style={rectangle, draw, thick, minimum width=3cm, minimum height=1.5cm, align=center},
    arrow/.style={->, >=stealth, thick, shorten <=1pt, shorten >=1pt}
]

\node[block] (seed) {$\bar{J} = \dfrac{d J}{d J} = 1$ \\ \footnotesize (Initial seed)};
\node[block, right=of seed] (xadj) {$\bar{x} = \bar{J} \cdot \dfrac{\partial J}{\partial x}$ \\ \footnotesize (SU2 adjoint)};
\node[block, right=of xadj] (zadj) {$\bar{z} = \bar{x} \cdot \dfrac{\partial x}{\partial z} = \dfrac{d J}{d z}$ \\ \footnotesize (Diffusion backprop)};

\draw[arrow] (seed.east) -- node[above, midway] {\footnotesize Backprop to $x$} (xadj.west);
\draw[arrow] (xadj.east) -- node[above, midway] {\footnotesize Backprop to $z$} (zadj.west);

\end{tikzpicture}
\caption{Reverse-mode automatic differentiation (AD) flow for computing $\frac{dJ}{dz}$. The computation begins with the initial seed $\bar{J}=1$, propagates backward through the SU2 adjoint to obtain $\bar{x}$, then through the diffusion model to obtain $\bar{z}$, which equals the final gradient $\frac{dJ}{dz}$.}
\label{fig:rev_ad_flow}
\end{figure}

Since SU2 applies reverse-mode AD to compute the discrete adjoint $\frac{\partial J}{\partial x}$ \cite{albring2016efficient}, and because a diffusion model \textemdash realized based on an artificial neural network  \textemdash naturally supports backpropagation, the vector Jacobian product in \eqref{eq:adjoint_dm} establishes a fully differentiable pipeline through reverse-mode AD (backpropagation). This computational flow, illustrated in Figure \ref{fig:rev_ad_flow}, consists of chaining two reverse-mode AD operations:

\begin{itemize}
    \item [1.] \textit{SU2 discrete adjoint:} Starting from the initial seed $\bar{J} = dJ/dJ = 1$, SU2 computes the adjoint $\bar{x} = \bar{J} \cdot \partial J/\partial x$ through its discrete adjoint implementation that is based on reverse-mode AD .
    
    \item [2.] \textit{Diffusion model backpropagation:} The output adjoint $\bar{x}$ from SU2 becomes the input seed for the diffusion model's backward pass, which computes the final gradient $\bar{z} = \bar{x} \cdot \partial x/\partial z = dJ/dz$ through neural network backpropagation.
\end{itemize}


\subsubsection{Backpropagation through Diffusion Model}

The computation of $\bar{x} \cdot \frac{\partial x}{\partial z}$ is realized by backpropagation through the diffusion sampling process, which leverages two fundamental properties:

\begin{enumerate}
    \item [1.] \textit{Iterative structure of diffusion models}: Diffusion models generate samples through sequential evaluations of a single neural network (typically a U-Net) across $T$ timesteps. As shown in \eqref{eq:ddim_sampling}, the forward (generative) process can be expressed as:
    \begin{equation*}
        x_{t-1} = \mathcal{D}_t(x_t) \quad (t=T,T-1,\dots,1),
    \end{equation*}
    where $\mathcal{D}_t$ represents the denoising network at time step $t$. The same network is applied at each step, differing only in the timestep conditioning.
    
    \item [2.] \textit{Differentiation of neural networks}: Backpropagation through a single network evaluation $\mathcal{D}_t$ to its input variable is a foundational operation in deep learning, which is of particular importance to Scientific Machine Learning. For example, it is extensively used in physics-informed neural networks (PINNs) \cite{raissi2019physics}, where output physical quantities (e.g., velocity $u$, pressure $p$) are differentiated with respect to input coordinates (e.g., spatial and temporal coordinates $x$, $t$) to formulate and enforce PDE constraints.
\end{enumerate}

The total backward pass of $\bar{x} \cdot \frac{\partial x}{\partial z}$ is therefore established by chaining the single backward passes of the U-net through all timesteps:
\begin{equation}
    \bar{x} \cdot \frac{\partial x_0}{\partial x_T} = \bar{x} \cdot \prod_{t=1}^{T} \frac{\partial x_{t-1}}{\partial x_t} = \bar{x} \cdot \prod_{t=1}^{T} \frac{\partial \mathcal{D}_t(x_t)}{\partial x_t},
    \label{eq:diffusion_chain}
\end{equation}
where $x_0 = x$ is the generated output parameter and $x_T = z$ is the latent variable. Since each $\mathcal{D}_t$ is differentiable and the composition is a finite sequence, the entire backward pass can be efficiently computed through reverse-mode automatic differentiation. 

\subsubsection{Integration into Existing Adjoint-Based Optimization Frameworks}

The manifold-constrained framework \eqref{eq:mcaso} provides distinct implementation benefits. First, it maintains compatibility with arbitrary existing shape parameterizations by enforcing manifold constraints rather than replacing parameterization methods. This is implemented through reverse-mode automatic differentiation, which propagates gradients from existing adjoint solvers through the diffusion model's generative process (Fig. 2). Second, the diffusion model trains exclusively offline, limiting computational overhead to inference during optimization \textemdash establishing it as an \textit{inference-time algorithm}. The fidelity of the learned manifold scales with training data quality and quantity, while specialized models for distinct problem classes (e.g., subsonic vs. transonic flows) can be deployed interchangeably without framework modifications. 


\section{Generative Modeling of Airfoil Shape Manifold with Diffusion Model}
\label{sec:generative_modeling}

\subsection{Diffusion Model Training}

The training data of our diffusion model is based on the airfoil dataset from the UIUC airfoil database. Since these airfoil data are represented using (x,y) coordinates, we have preprocessed the data to obtain their respective Hicks-Henne parameterizations. For each airfoil, we formulate and solve the inverse optimization problem:
\begin{equation}
    \min_{\mathbf{p} \in \mathbb{R}^{40}} \sum_{i=1}^{200} \left[ \left(y^\text{up}_i(\mathbf{p}) - y^{\text{target,up}}_i\right)^2 + \left(y^\text{lo}_i(\mathbf{p}) - y^{\text{target,lo}}_i\right)^2 \right]
\end{equation}  
where $\mathbf{p} = [p_1^{\text{up}}, ..., p_{20}^{\text{up}}, p_1^{\text{lo}}, ..., p_{20}^{\text{lo}}]^T$ contains the Hicks-Henne parameters for upper/lower surfaces. Note, we have first resampled the airfoil coordinates to 200 interpolated points on upper/lower surfaces to faciliate a more robust inverse problem solving. The preprocessed Hickse-Henne parameters were then normalized from 0 to 1 for our diffusion model training. In total, our training dataset consists of 1568 airfoils represented with Hicks-Henne parameters.

We train a diffusion model that generates 1D sequence for the Hicks-Henne parameters instead of the common 2D sequence used for image generations. To this purpose, we make use of a 1D denoising diffusion probabilistic model using the \texttt{GaussianDiffusion1D} framework from the \href{https://github.com/lucidrains/denoising-diffusion-pytorch}{lucidrains/denoising-diffusion-pytorch} repository \cite{ho2020denoising}. Thereby, only a few adaptations were made to accommodate the shape and size of our training data:
\begin{itemize}
    \item The \texttt{GaussianDiffusion1D} module is initialized with \texttt{seq\_length = 40} to align with the dimensionality of our airfoil representations. 
    \item  We employ \texttt{Unet1D} with input \texttt{channels = 1} instead of the default 32, matching the scalar nature of each element in our 40-dimensional HHM vectors.
    \item To accommodate our limited dataset ($N = 1,568$ samples), we made adjustment in the hyperparameters:
    \begin{itemize}
        \item Learning rate: \texttt{train\_lr = 1e-5} (default 8e-5)
        \item Batch size: \texttt{train\_batch\_size = 2} (default 32)
        \item Training steps: \texttt{train\_num\_steps = 140,000} (5$\times$ reduction from the default 700,000)
    \end{itemize}
    \item We utilize Denoising Diffusion Implicit Models (DDIM) for efficient and deterministic generation.
\end{itemize}

\subsection{Learned Airfoil Parameter Manifold of Diffusion Model}
\label{sec:learned_manifold}

Our central hypotheses are (A1) aerodynamically viable shapes \emph{locally} form low-dimensional manifolds in the Hicks–Henne parameter space and (A2) a diffusion model with strong generalization should recover the score function of the underlying \emph{population distribution} perfectly.
Under these hypotheses, the adjoint-based optimization in \Cref{sec:method} enables the discovery of novel airfoils.
In this section, we first derive the theoretical signatures that must appear if the above two hypotheses hold. 
We then verify these signatures empirically, providing empirical evidence that the learned manifold constraint is active and that its particular formulation underpins our success.

Formally, the above two assumptions are stated as follows:
\begin{itemize}
\item[(A1)] [Local manifold structure] Let $\mu$ be the population distribution that generates the empiricial distribution $p_{\mathrm{data}}$. 
Let $\tilde x$ be a point in $\mathrm{supp}(\mu)$. There exists some open neighborhood $\mathrm{N}(\tilde x)$ of $\tilde x$ such that $\cM = \mathrm{supp}(\mu) \cap \mathrm{N}(\tilde x)$ is a compact, smooth, embedding submanifold without boundary. Let $k < d$ be the dimension of $\cM$.
\item[(A2)] [Perfect score matching] The diffusion model perfectly learns the population denoiser $\epsilon_\theta$, achieving zero loss in (\ref{eqn_obj_DDPM}) for $x_0 \sim \mu$.
\end{itemize}
\paragraph{Predicted observations.} Under these two assumptions, we state the predicted observation as follows.
\begin{prediction*}[Low-rank Jacobian]
    Suppose that the above two assumptions hold.
    If the manifold constraint is active at $\tilde x$, the Jacobian of the score function of the diffusion model should be low-rank at $\tilde x$.
\end{prediction*}
\noindent The above prediction can be derived from the following three steps.
\begin{itemize}
    \item Let $X_0$ be a random variable distributed according to $\mu$.
    Let $X_1$ be the random variable after taking one step of the forward diffusion process in DDPM, i.e. 
    \begin{equation*}
        X_1 = \sqrt{1-\beta_1} X_0 + \sqrt{\beta_1} \xi,
    \end{equation*}
    where $\xi$ is some standard Gaussian random vector with matching dimension. 
    The distribution of $X_1$, denoted as $p_{\beta_1}$ can be identified as convolving the $\mu$ with a transition kernel $K_{\beta_1} = \cN(y; \sqrt{1-\beta_1}x, \beta_1 I)$, i.e.
    \begin{equation} 
        p_{\beta_1}(y) = \frac{1}{Z_{\beta_1}} \int \exp(-\frac{\|y - \sqrt{1-\beta_1}x\|^2}{2\beta_1}) \rmd \mu(x) = K_{\beta_1} \ast \mu.
    \end{equation}
    Here $\ast$ denotes convolution of measures and $Z_{\beta_1}$ is some normalizing factor.
    \item 
    With the reparameterization $s_\theta = -\epsilon_\theta / \beta_1$, $s_\theta$ exactly matches $\nabla \log p_{\beta_1}$, the score function of $p_{\beta_1}$.
    \item Define $V_{\beta_1}(x) = {\beta_1} \log K_{\beta_1} \ast \mu$.
    We have that for all $x \in \mathrm{N}(\tilde x)$, $V_{\beta_1}$ is twice continuously differentiable at $x$ and 
    \begin{equation}
        \lim_{{\beta_1}\rightarrow 0} V_{\beta_1}(x) = \frac{1}{2}\mathrm{dist}^2(x, \cM).
    \end{equation}
    Moreover, the matrix of $I_d - \nabla^2 V_{\beta_1}(\tilde x)$ matches the orthogonal projection matrix of the tangent space at $\tilde x$, i.e.
    \begin{equation}
        I_d - \lim_{{\beta_1}\rightarrow 0} \nabla^2 V_{\beta_1}(\tilde x) = \cP(\cT_{\tilde x}\cM).
    \end{equation}
    Recall that for a $k$ dimensional subspace embeded in a $d$-dimensional ambient space, its unique (positive semi-definite) orthogonal projection matrix has $k$ eigenvalues that are all equal to $1$ and the rest $d-k$ eigenvalues that are all equal to $0$.
    Consequently, if $\cM$ is a $k$ dimensional manifold, we have $\mathrm{rank}(\nabla^2 V_0(\tilde x)) = d - k$.
    By the continuity of the Hessian $\nabla^2 V_{\beta_1}$, the Jacobian matrix of $\beta_1 \cdot s_\theta(\tilde x)$, which exactly matches $\nabla^2 V_{\beta_1}(\tilde x)$, should be approximately low-rank for every sufficiently small $\beta_1$.
\end{itemize}

\paragraph{Empricial evidence.}
We now provide empirical evidence to support our hypothesis that the manifold constraint is active.
We highlight that our empirical observations are \emph{necessary} if the manifold constraint is active and are \emph{not sufficient} to establish the validity of the assumptions (A1) and (A2).

\begin{figure}[h]
    \centering
    \includegraphics[width=0.5\linewidth]{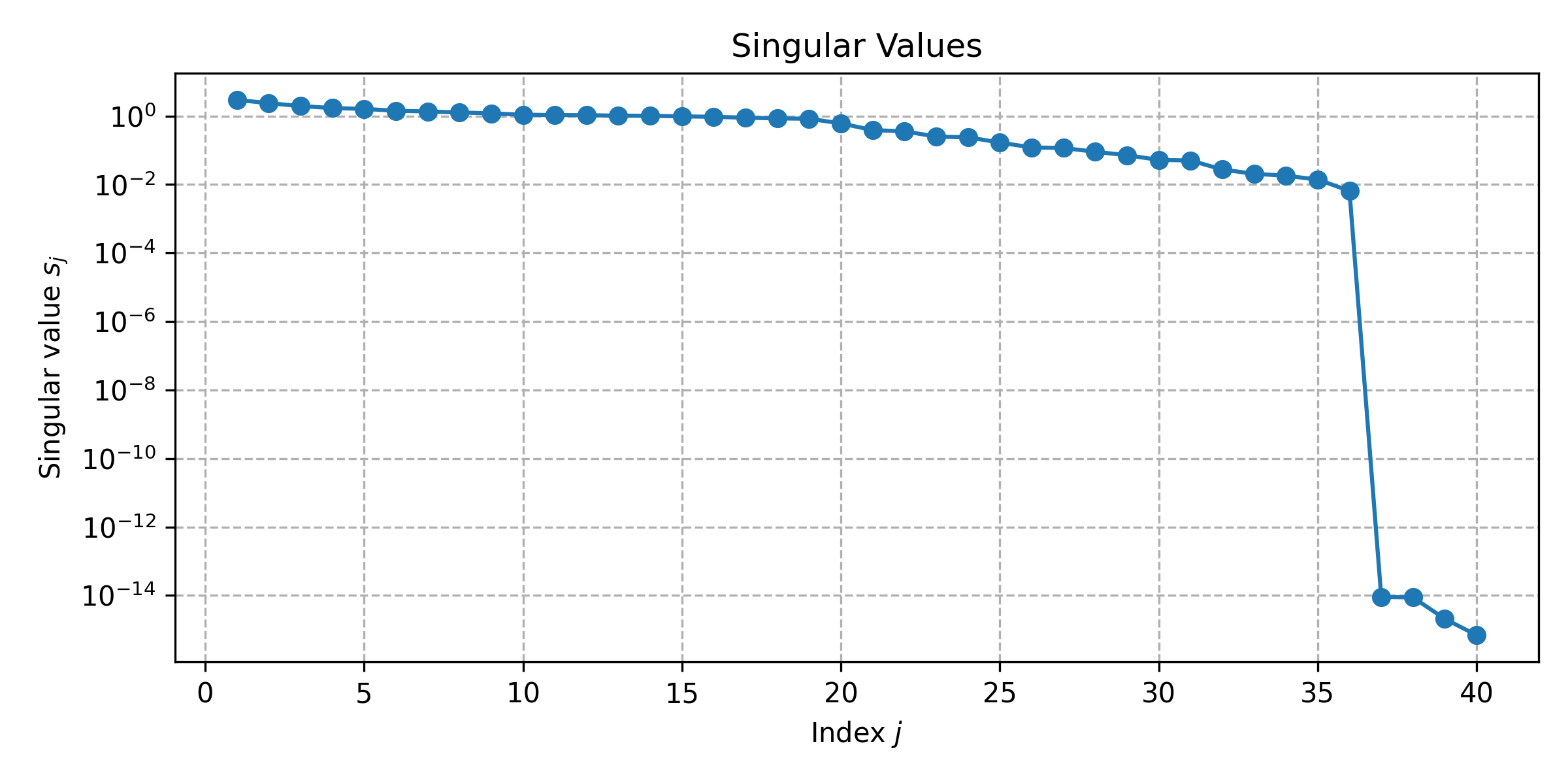}
    \caption{Singular values of the Jacobian of the denoiser at step $t = 1$ for optimized shape with and constraint $C_l \geq 0.3$ and thickness to chord ratio $ t_c \geq 0.12$. }
    \label{fig:singular_values_model_opt3_log}
\end{figure}
\begin{figure}[h]
    \centering
    \includegraphics[width=0.5\linewidth]{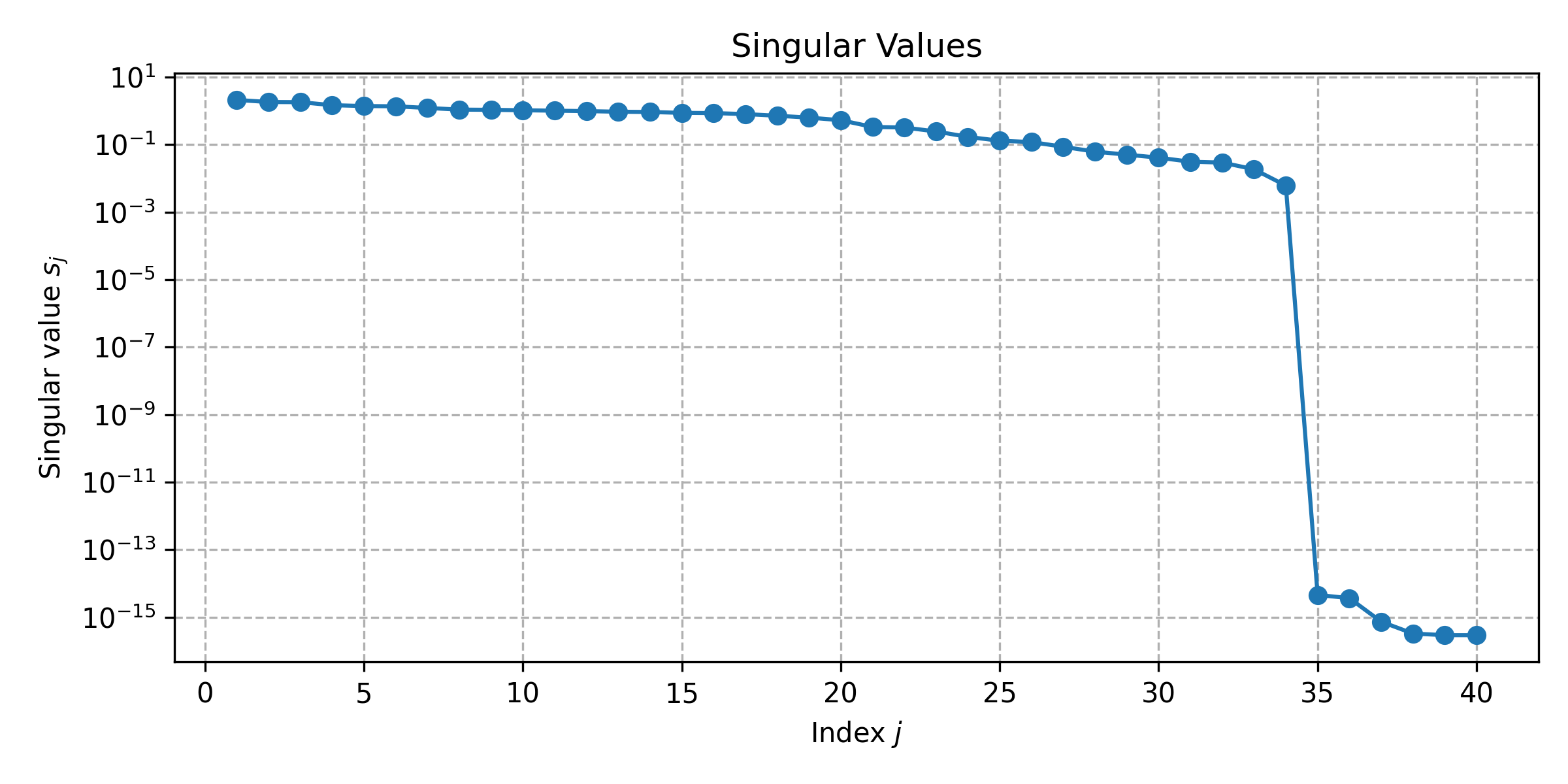}
    \caption{Singular values of the Jacobian of the denoiser at step $t = 1$ for optimized shape with and constraint $C_l \geq 0.4$ and thickness to chord ratio $ t_c \geq 0.12$. }
    \label{fig:singular_values_model_opt2_log}
\end{figure}
\begin{figure}[h]
    \centering
    \includegraphics[width=0.5\linewidth]{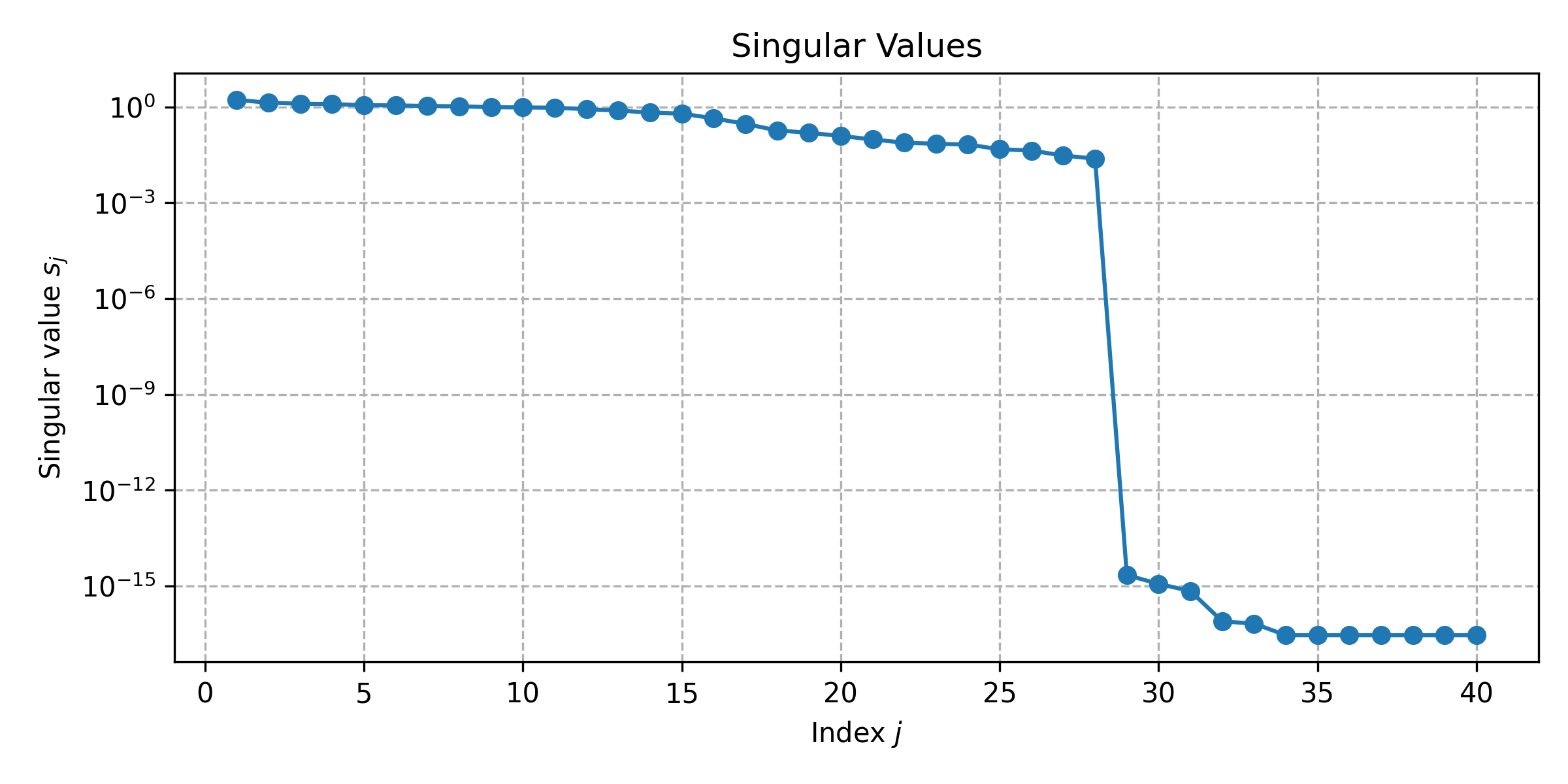}
    \caption{Singular values of the Jacobian of the denoiser at step $t = 1$ for optimized shape with and constraint $C_l \geq 0.5$ and thickness to chord ratio $ t_c \geq 0.12$. }
    \label{fig:singular_values_model_opt1_log}
\end{figure}

We analyze the learned manifold of our diffusion model by examining the singular values of the Jacobian of the score function at sampling time $t = 1$. In Figures \ref{fig:singular_values_model_opt3_log} - \ref{fig:singular_values_model_opt1_log}, we plot the singular values for three optimized designs, each obtained by solving a different optimization problem (see the next section). For the purpose of the study in this section, it suffices to note that these designs correspond to different lift constraint values $C_L = 0.3, 0.4, 0.5$. We observe that the singular values of the Jacobian matrix of $\beta_1 \cdot s_\theta(\tilde x)$ all have drastic drops at some index $d-k$ for $k > 1$ ($k = 4, 5, 11$ respectively), matching the low-rank prediction stated above. Notably, the effective rank decreases as the lift constraint becomes more restrictive.

\section{Optimization Results}
\label{sec:results}

We apply our proposed method to aerodynamic shape optimization, where the diffusion model generates airfoil geometries parameterized by Hicks-Henne variables $x \in \mathbb{R}^d$ from latent variables $z \in \mathcal{Z}$. The Hicks-Henne variables define a surface deformation based on the NACA0012 shape as reference configuration.

We consider the constrained optimization problem under transonic flow conditions at Mach number $0.8$ and angle of attack $2.31^\circ$. The optimization problem reads
\begin{equation}
    \begin{aligned}
    \min_{x \in \mathcal{X}_G} \quad & C_d(x) \\
    \text{s.t.} \quad & C_l(x) \geq \underline{C_l}, \\
                      & t_c(x) \geq \underline{t_c}, \\
    \end{aligned}
    \label{eq:data-informed_shape_design}
\end{equation}
where $t_c$ denotes the thickness to chord ratio of the airfoil. For a thorough assessment and comparison of our proposed approach with conventional approach, we conduct experiments on a series of design optimization problems, with $\underline{C_l} \times \underline{t_c} = \{0.3, 0.4, 0.5 \} \times \{ 0.105, 0.12 \}$. Furthermore, we use RAE2822 and NACA0012 as different initializations $x_0$ and employ two well-established nonlinear optimization solver SLSQP \cite{kraft1988software} and IPOPT \cite{wachter2006implementation}.

We perform RANS simulations with the Spalart-Allmaras turbulence model. We use 40 Hicks-Henne parameters, and hence 40 latent space parameters for diffusion model based optimization. Both the forward and adjoint runs are performed using SU2. The experiments were run on a cluster, where we use a GPU for the diffusion model forward and backward process, and at each optimization iteration, CPU nodes are allocated for the different SU2 runs.


We choose the best optimal solution $x^\star$ with constraint violations $\varepsilon_{rel} \lessapprox 2e-3$ with
\begin{equation}
\varepsilon_{rel} = \max\left(\frac{C_l(x^\star)- \underline{C_l}}{\underline{C_l}}, \frac{t_c(x^\star) - \underline{t_c}}{\underline{t_c}}\right).
\end{equation}

The results are documented in tables, in which each row documents a constrained drag-minimization problem with specified lower bounds on lift coefficient $\underline{C_l}$ and thickness $\underline{t_c}$. From left to right, the columns report the number of objective evaluations ($\# J$), the number of gradient evaluations ($\# \nabla J)$, the achieved lift coefficient $C_l$ and thickness $t_c$, and the optimied drag count $C_d$. We should emphasize that in all of our experiments, no scaling (fine-tuning) parameter were used for our proposed method, while a careful parameter study were performed to determine a usable combination of scaling parameters for the Hicks-Henne method.

\subsection{Results with RAE2822 Initialization and SLSQP Optimizer}
\label{sec:slsqp_rae}

In Table \ref{tab:rae_slsqp} we compare the performance of our proposed method with the Hicks-Henne parameterization, using the RAE2822 airfoil as the initial geometry and SLSQP as the optimizer. For the moderate constraints $(\underline{C_l}, \underline{t_c}) = (0.30, 0.105)$, $(\underline{C_l}, \underline{t_c}) = (0.30, 0.120)$, and $(\underline{C_l}, \underline{t_c}) = (0.40, 0.105)$, DM outperforms HHM with marginally lower drag counts. As the lift and thickness bounds increases, the advantage of DM grows. Under the most difficult constraints $(0.50, 0.105)$ and $(0.50, 0.120)$, HHM not only results in significanly higher drag counts but also suffers from constraint violations (market with asterisks). Finally, it is evident that DM constently requires fewer function and gradient evaluations than HHM except the last test case with $(\underline{C_l}, \underline{t_c}) = (0.50, 0.120)$, where the HHM results in a design that is substantially worse than DM. 

\begin{table}[ht]
  \centering
  \caption{Comparison of Method DM vs.\ HHM with RAE2822 initialization and SLSQP Optimizer}
  \label{tab:rae_slsqp}
  \begin{tabular}{|cc|cc|cc|cc|cc|cc|}
    \hline
    \multicolumn{2}{|c|}{Problem}
      & \multicolumn{2}{c|}{\# $J$}
      & \multicolumn{2}{c|}{\# $\nabla J$}
      & \multicolumn{2}{c|}{$C_l$}
      & \multicolumn{2}{c|}{$t_c$}
      & \multicolumn{2}{c|}{$C_d$ (count)} \\
    \cline{1-2} \cline{3-4} \cline{5-6} \cline{7-8} \cline{9-10} \cline{11-12}
      $\underline{C_l}$ & $\underline{t_c}$
      & DM & HHM
      & DM & HHM
      & DM & HHM
      & DM & HHM
      & DM & HHM \\
    \hline
    0.30 & 0.105 &  10  &  14  &  10  &  14  &  0.3001  &  0.2999  &  0.1049  &  0.1049  &    218.7 &  221.0  \\
    0.30 & 0.120 &  15  &  16  &  14  &  16  &  0.2998  &  0.2999  &  0.1199  &  0.1197  &  316.9  &  326.6  \\
    0.40 & 0.105 &  13  &  21  &  13  &  18  &  0.4000  &  0.3999  &  0.1049  &  0.1049  &  286.7  &   286.8 \\
    0.40 & 0.120 &  13  &  21  &  11  &  20  &  0.3995  &  0.3993  &  0.1199  &  0.1195$^*$  &   411.2 &  440.6  \\
    0.50 & 0.105 &  18  &  21  &  16  &  19  & 0.4999    &  0.4969$^*$ &  0.1050   & 0.1042$^*$  &   392.3 &  413.9  \\
    0.50 & 0.120 &  28  &  23   &  16  &  7  &  0.5002  & 0.4902$^*$   &  0.1197  & 0.1127$^*$   & 560.8    & 921.7   \\
    \hline
  \end{tabular}
\end{table}

Figure \ref{fig:slsqp_rae_t1} and \ref{fig:slsqp_rae_t2} illustrate the optimized airfoil shapes and their respective CFD solutions of both methods. DM consistently produces airfoils with supercritical characteristics, whereas HHM geometries degrade as the optimization problem difficulty increases, indicating the optimizer is trapped into inferior local minima. Even in the simpler cases where both methods find solutions that satisfy the constraints, DM converges to subtly different but aerodynamically superior local solutions compared to HHM.

\newlength{\mysubfigwidth}
\setlength{\mysubfigwidth}{0.45\textwidth}  

\begin{figure}[htp]
  \centering

  \begin{subfigure}[b]{\textwidth}
    \centering
    \begin{subfigure}[b]{\mysubfigwidth}
      \includegraphics[width=\linewidth]{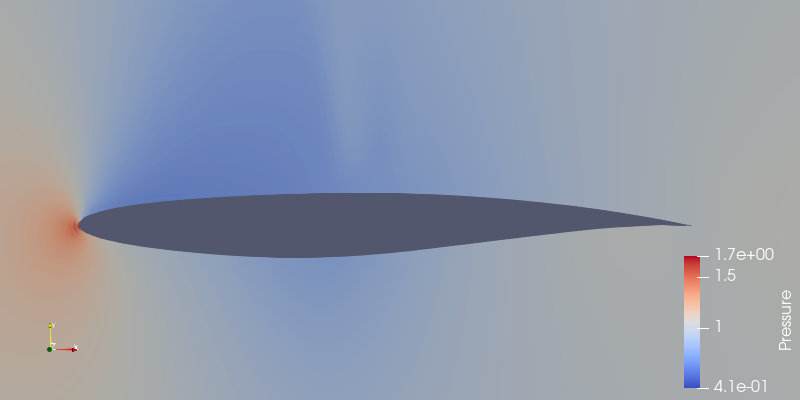}
    \end{subfigure}\hfill
    \begin{subfigure}[b]{\mysubfigwidth}
      \includegraphics[width=\linewidth]{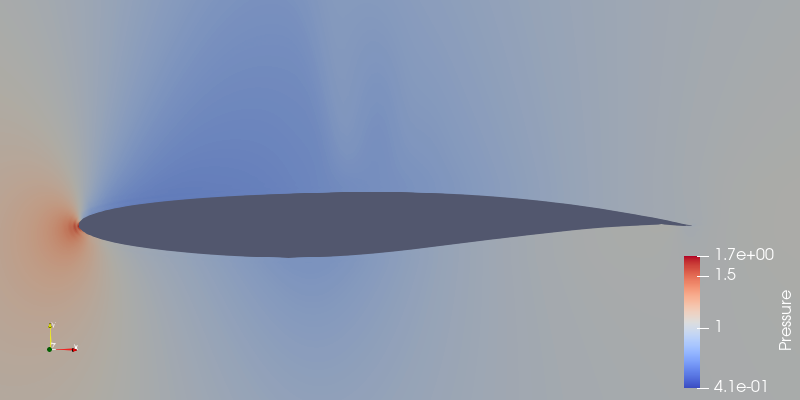}
    \end{subfigure}
    \subcaption{Bounds $(\underline{C_l}, \underline{t_c}) = (0.30,\,0.105)$}
  \end{subfigure}
  \medskip

  \begin{subfigure}[b]{\textwidth}
    \centering
    \begin{subfigure}[b]{\mysubfigwidth}
      \includegraphics[width=\linewidth]{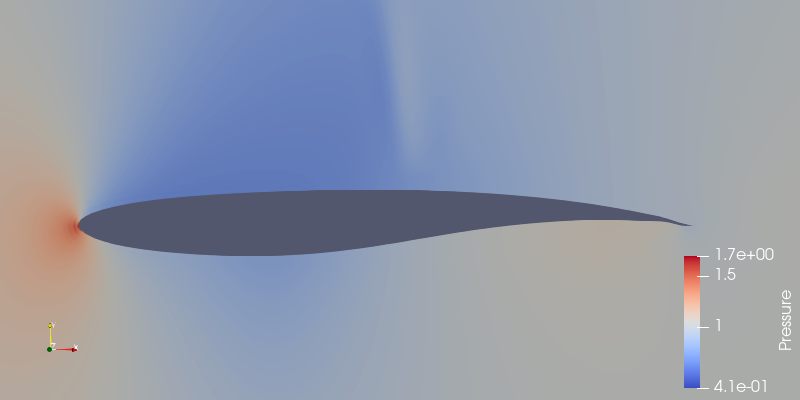}
    \end{subfigure}\hfill
    \begin{subfigure}[b]{\mysubfigwidth}
      \includegraphics[width=\linewidth]{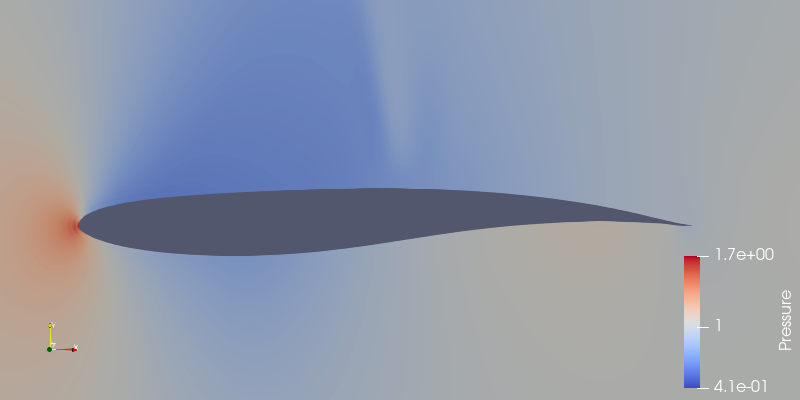}
    \end{subfigure}
    \subcaption{Bounds $(\underline{C_l}, \underline{t_c}) = (0.40,\,0.105)$}
  \end{subfigure}
  \medskip

  \begin{subfigure}[b]{\textwidth}
    \centering
    \begin{subfigure}[b]{\mysubfigwidth}
      \includegraphics[width=\linewidth]{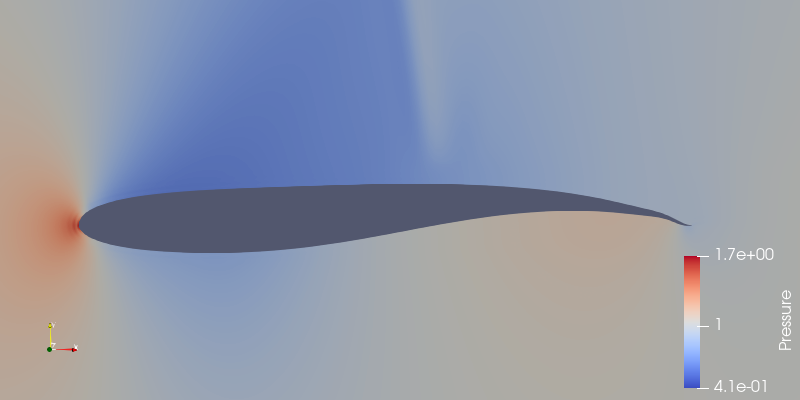}
    \end{subfigure} \hfill
    \begin{subfigure}[b]{\mysubfigwidth}
      \includegraphics[width=\linewidth]{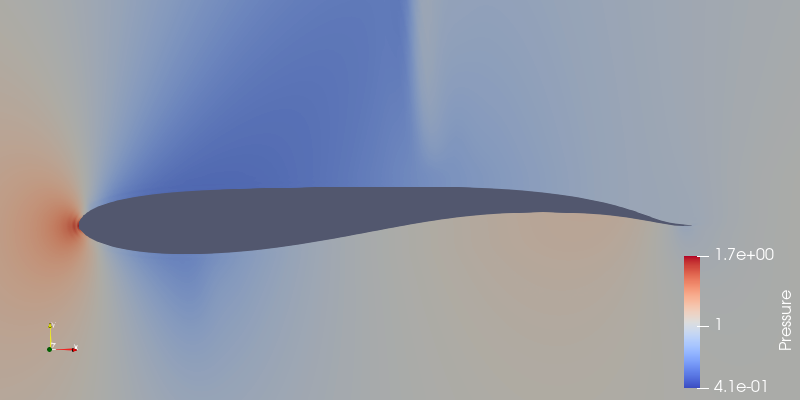}
    \end{subfigure}
    \subcaption{Bounds $(\underline{C_l}, \underline{t_c}) = (0.50,\,0.105)$}
  \end{subfigure}
  \smallskip

  \caption{SLSQP optimized shape and flow results (pressure) for different lower constraint bounds $(\underline{C_l}, \underline{t_c}) \in \{0.3, 0.4, 0.5\} \times \{0.105\}$, comparing the DM- and HHM-based methods (left/right in each row). Initialized with RAE2822.}
  \label{fig:slsqp_rae_t1}
\end{figure}

\begin{figure}[htp]
  \centering

  \begin{subfigure}[b]{\textwidth}
    \centering
    \begin{subfigure}[b]{\mysubfigwidth}
      \includegraphics[width=\linewidth]{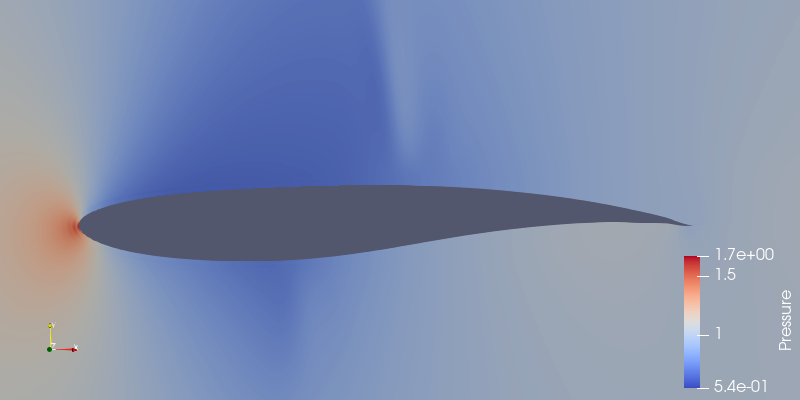}
    \end{subfigure}\hfill
    \begin{subfigure}[b]{\mysubfigwidth}
      \includegraphics[width=\linewidth]{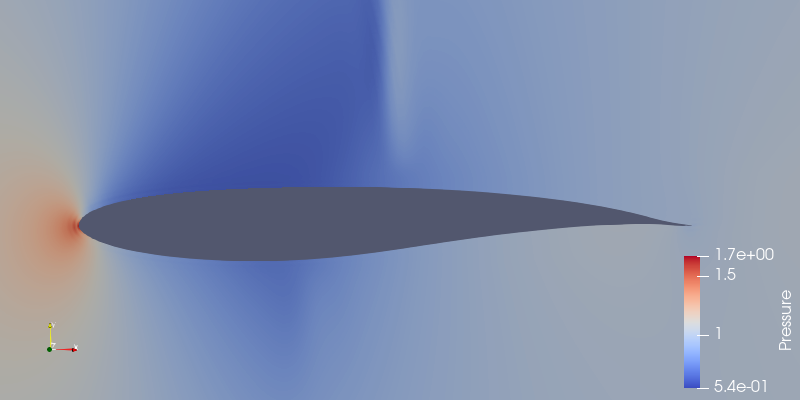}
    \end{subfigure}
    \subcaption{Bounds $(\underline{C_l}, \underline{t_c}) = (0.30,\,0.120)$}
  \end{subfigure}
  \medskip

  \begin{subfigure}[b]{\textwidth}
    \centering
    \begin{subfigure}[b]{\mysubfigwidth}
      \includegraphics[width=\linewidth]{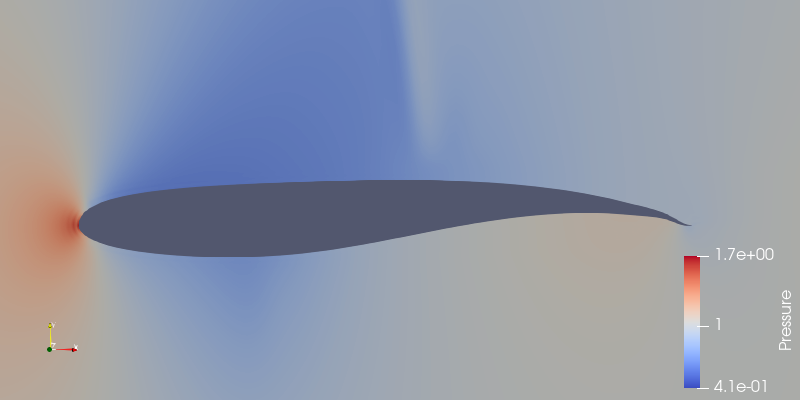}
    \end{subfigure}\hfill
    \begin{subfigure}[b]{\mysubfigwidth}
      \includegraphics[width=\linewidth]{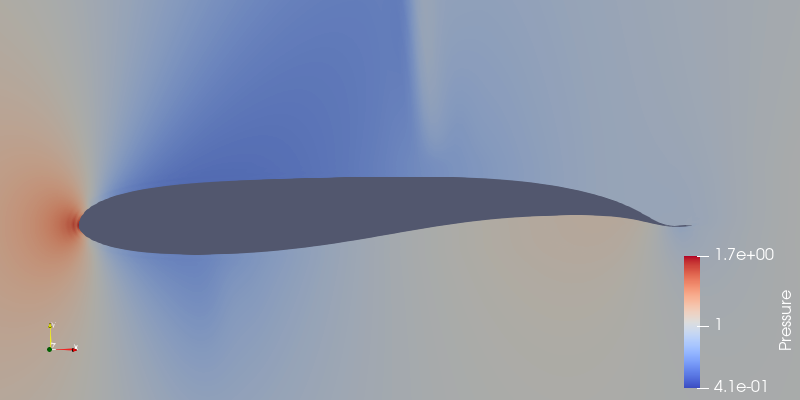}
    \end{subfigure}
    \subcaption{Bounds $(\underline{C_l}, \underline{t_c}) = (0.40,\,0.120)$}
  \end{subfigure}
  \medskip

  \begin{subfigure}[b]{\textwidth}
    \centering
    \begin{subfigure}[b]{\mysubfigwidth}
      \includegraphics[width=\linewidth]{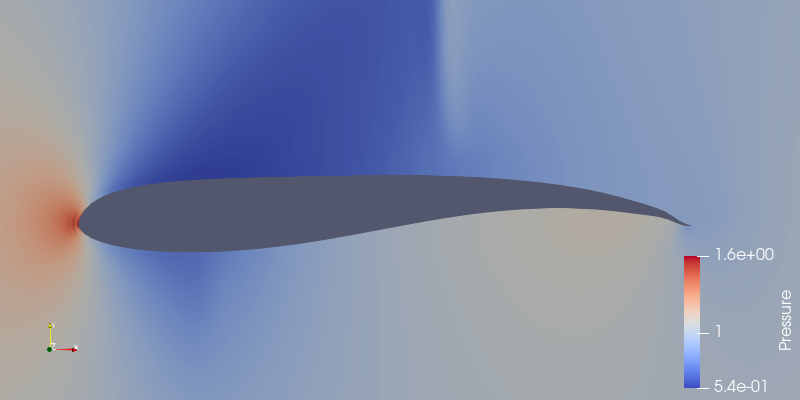}
    \end{subfigure}  \hfill
    \begin{subfigure}[b]{\mysubfigwidth}
      \includegraphics[width=\linewidth]{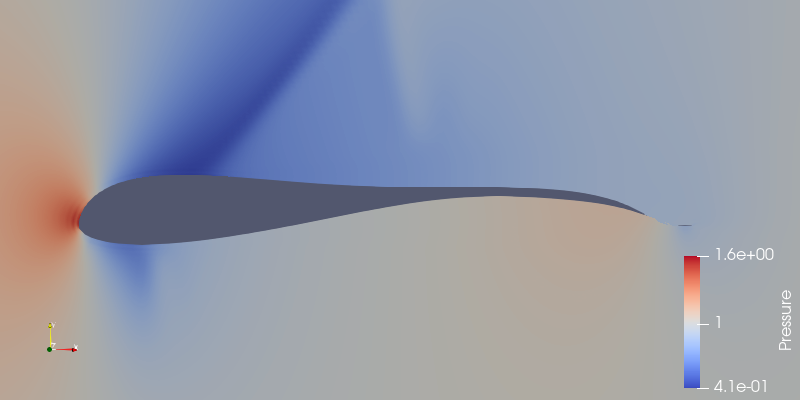}
    \end{subfigure}
    \subcaption{Bounds $(\underline{C_l}, \underline{t_c}) = (0.50,\,0.120)$}
  \end{subfigure}

  \caption{SLSQP optimized shape and flow results (pressure) for different lower constraint bounds $(\underline{C_l}, \underline{t_c}) \in \{0.3, 0.4, 0.5\} \times \{0.12\}$, comparing the DM- and HHM-based methods (left/right in each row). Initialized with RAE2822.}
  \label{fig:slsqp_rae_t2}
\end{figure}

\subsection{Results with NACA0012 Initialization and SLSQP Optimizer}
\label{sec:slsqp_naca}

In a second series of tests, we replace the RAE2822 airfoil with the NACA0012 airfoil as the initializing shape. Under the transonic flow conditions considered, the symmetry of NACA0012 places the initialization further away from the constrained-optimum in the optimization landscape than the cambered RAE2822. Table \ref{tab:naca_slsqp} summarizes the optimization results of both DM and HHM, again employing SLSQP as the optimization algorithm. Similar to previous test seires, DM outperforms HHM across all cases, and its advantage grows with problem difficulty. Comparing the two initializations, we observe the following:

\begin{itemize}
    \item [1] When initialized from NACA0012, which lies farther from the constrained-optimum, HHM results in noticeably infeior solutions than with RAE2822 initialization.
    \item [2] For half of the test problems, DM yields in even lower drag counts when initialized from NACA0012 than from RAE2822. We attribute this to the manifold constraint $x \in \mathcal{X}_G$, which simplifies the optimization landscape, and allows the optimizer to explore a broader region of the design space to navigate to superior local solutions.    
\end{itemize}

Figure \ref{fig:slsqp_naca_t1} - \ref{fig:slsqp_naca_t2} shows the optimized airfoil shape. We observe an up-tilting of the trailing edges of optimized shapes resulting from HHM. This phenomenon can be explained by examining the adjoint solutions, which exhibit concentrated high values on the suction side at the trailing edge under transonic flow conditions (see, e.g., Figure 6.4 in \cite{gauger2003adjungiertenverfahren}). This is physically perfectly valid for reducing the shock, but leads to suboptimal local solutions that trap the gradient-based optimization search. In the next section, we remedy this issue with physics insights by introducing scalings to the HHM parameters.

\begin{table}[H]
  \centering
  \caption{Comparison of Method DM vs.\ HHM with NACA0012 initialization and SLSQP Optimizer}
  \label{tab:naca_slsqp}
  \begin{tabular}{|cc|cc|cc|cc|cc|cc|}
    \hline
    \multicolumn{2}{|c|}{Problem}
      & \multicolumn{2}{c|}{\# $J$}
      & \multicolumn{2}{c|}{\# $\nabla J$}
      & \multicolumn{2}{c|}{$C_l$}
      & \multicolumn{2}{c|}{$t_c$}
      & \multicolumn{2}{c|}{$C_d$ (count)} \\
    \cline{1-2} \cline{3-4} \cline{5-6} \cline{7-8} \cline{9-10} \cline{11-12}
      $\underline{C_l}$ & $\underline{t_c}$
      & DM & HHM
      & DM & HHM
      & DM & HHM
      & DM & HHM
      & DM & HHM \\
    \hline
    0.30 & 0.105 & 37 & 19 & 33 & 17 & 0.3001 & 0.2996 & 0.1050 & 0.1048 & 214.6 & 245.5 \\
    0.30 & 0.120 & 12 & 19 & 11 & 19 &  0.3000 & 0.2999 & 0.1199 & 0.1198 &  311.7 & 362.0 \\
    0.40 & 0.105 & 10 & 20 & 9 & 19 & 0.4000 & 0.3999 & 0.1049 & 0.1049 & 296.5 & 361.3 \\
    0.40 & 0.120 & 24 & 17 & 16 & 8 & 0.4002 & 0.4088 & 0.1199 & 0.1161$^*$ & 402.9 & 863.2 \\
    0.50 & 0.105 & 10 & 14 & 8 & 5 & 0.4997 & 0.5069 & 0.1049 & 0.1033$^*$ & 388.9 & 804.1 \\
    0.50 & 0.120 & 62 & 19 & 36 & 6 & 0.4998 & 0.4217$^*$ & 0.1198 & 0.1116$^*$ & 576.7 & 835.2 \\
    \hline
  \end{tabular}
\end{table}




\setlength{\mysubfigwidth}{0.45\textwidth}

\begin{figure}[htp]
  \centering
  \begin{subfigure}[b]{\textwidth}
    \centering
    \begin{subfigure}[b]{\mysubfigwidth}
      \includegraphics[width=\linewidth]{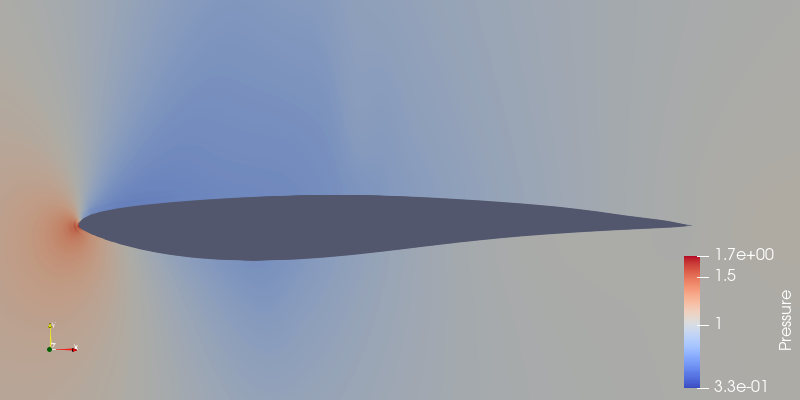}
    \end{subfigure}\hfill
    \begin{subfigure}[b]{\mysubfigwidth}
      \includegraphics[width=\linewidth]{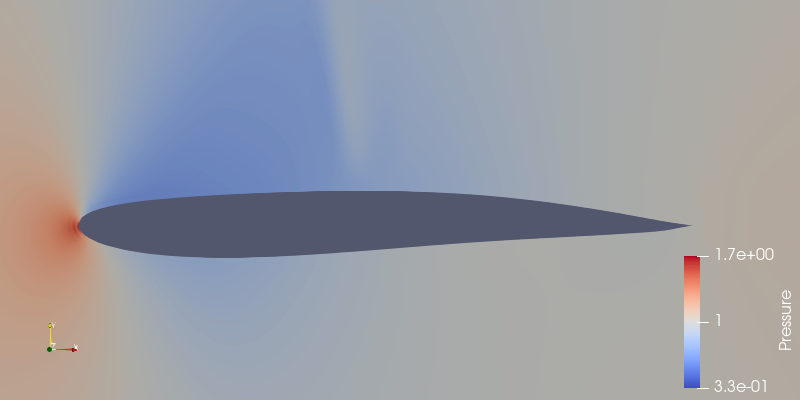}
    \end{subfigure}
    \subcaption{Bounds $(\underline{C_l}, \underline{t_c}) = (0.30,\,0.105)$}
  \end{subfigure}
  \medskip

  \begin{subfigure}[b]{\textwidth}
    \centering
    \begin{subfigure}[b]{\mysubfigwidth}
      \includegraphics[width=\linewidth]{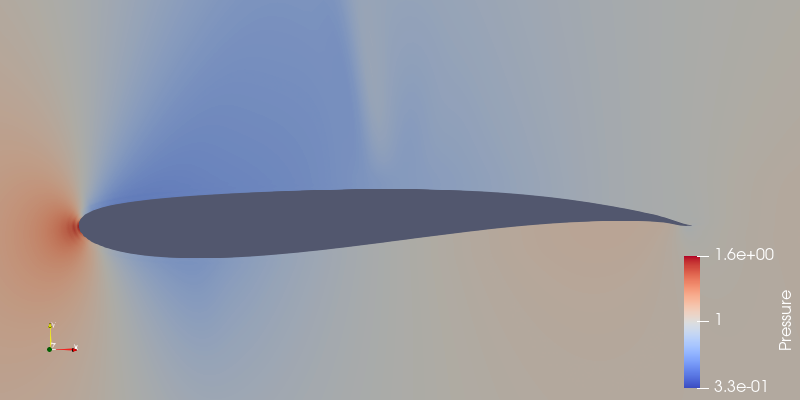}
    \end{subfigure}\hfill
    \begin{subfigure}[b]{\mysubfigwidth}
      \includegraphics[width=\linewidth]{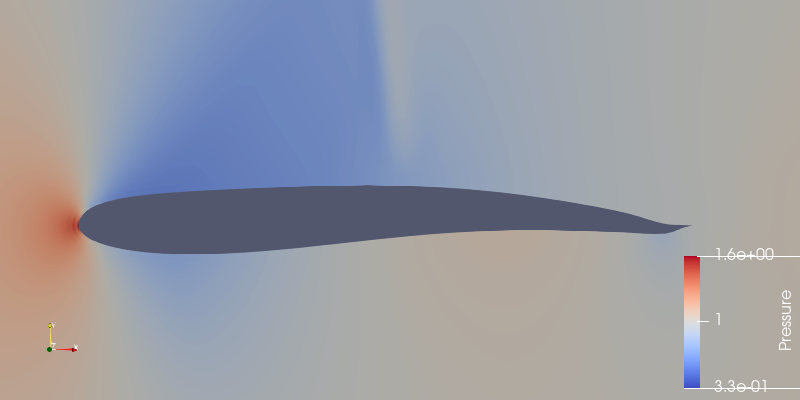}
    \end{subfigure}
    \subcaption{Bounds $(\underline{C_l}, \underline{t_c}) = (0.40,\,0.105)$}
  \end{subfigure}
  \medskip

  \begin{subfigure}[b]{\textwidth}
    \centering
    \begin{subfigure}[b]{\mysubfigwidth}
      \includegraphics[width=\linewidth]{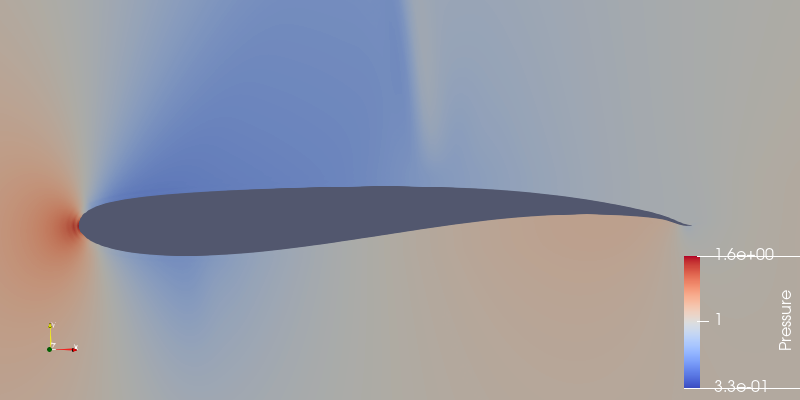}
    \end{subfigure}\hfill
    \begin{subfigure}[b]{\mysubfigwidth}
      \includegraphics[width=\linewidth]{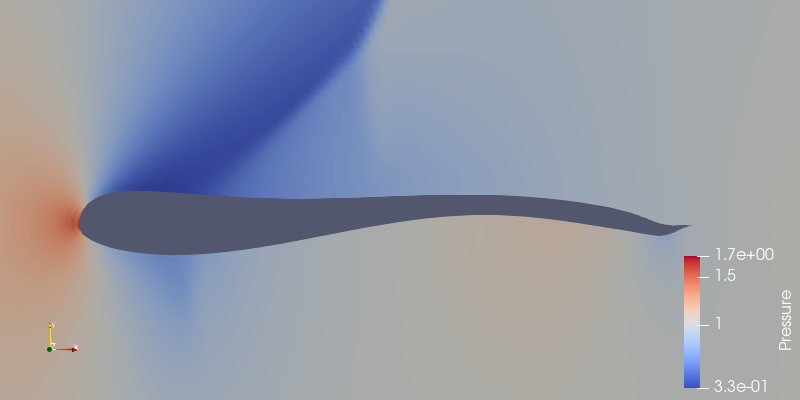}
    \end{subfigure}
    \subcaption{Bounds $(\underline{C_l}, \underline{t_c}) = (0.50,\,0.105)$}
  \end{subfigure}
  \smallskip

  \caption{SLSQP optimized shape and flow results (pressure) for different lower constraint bounds $(\underline{C_l}, \underline{t_c}) \in \{0.3, 0.4, 0.5\} \times \{0.105\}$, comparing DM vs.\ HHM (left/right). Initialized with NACA0012.}
  \label{fig:slsqp_naca_t1}
\end{figure}

\begin{figure}[htp]
  \centering

  \begin{subfigure}[b]{\textwidth}
    \centering
    \begin{subfigure}[b]{\mysubfigwidth}
      \includegraphics[width=\linewidth]{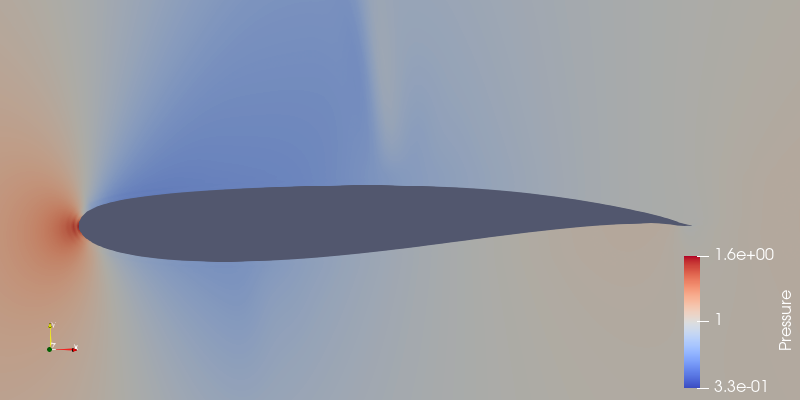}
    \end{subfigure}\hfill
    \begin{subfigure}[b]{\mysubfigwidth}
      \includegraphics[width=\linewidth]{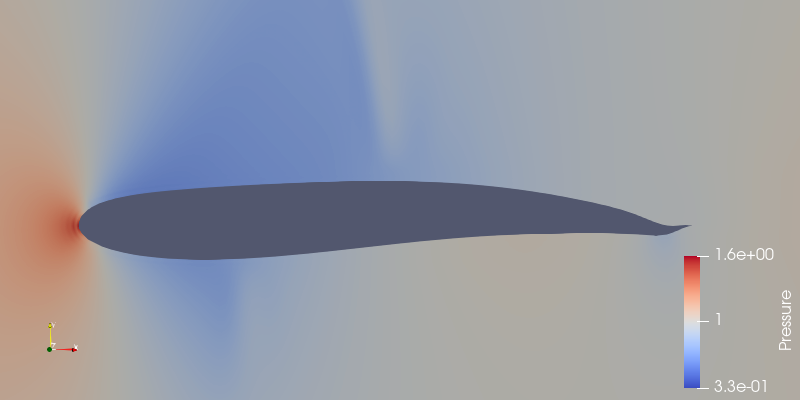}
    \end{subfigure}
    \subcaption{Bounds $(\underline{C_l}, \underline{t_c}) = (0.30,\,0.120)$}
  \end{subfigure}
  \medskip

  \begin{subfigure}[b]{\textwidth}
    \centering
    \begin{subfigure}[b]{\mysubfigwidth}
      \includegraphics[width=\linewidth]{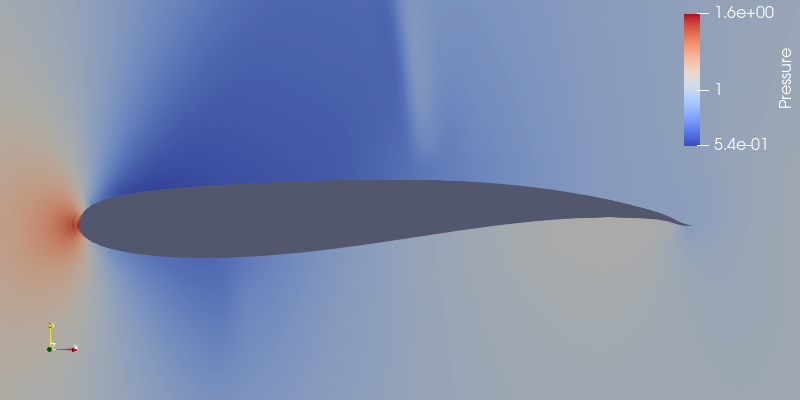}
    \end{subfigure}\hfill
    \begin{subfigure}[b]{\mysubfigwidth}
      \includegraphics[width=\linewidth]{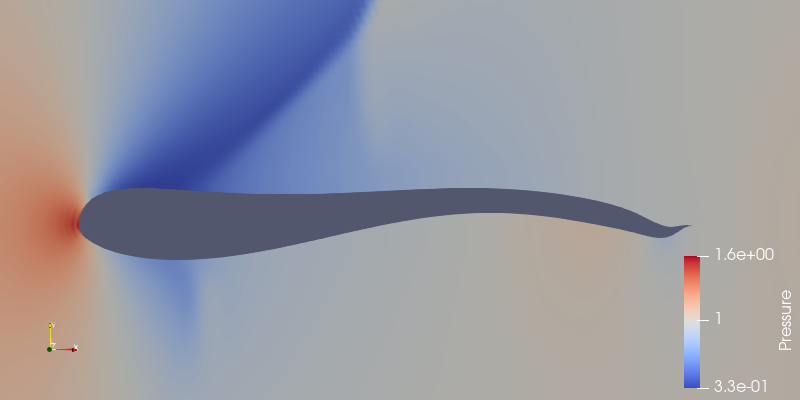}
    \end{subfigure}
    \subcaption{Bounds $(\underline{C_l}, \underline{t_c}) = (0.40,\,0.120)$}
  \end{subfigure}
  \medskip

  \begin{subfigure}[b]{\textwidth}
    \centering
    \begin{subfigure}[b]{\mysubfigwidth}
      \includegraphics[width=\linewidth]{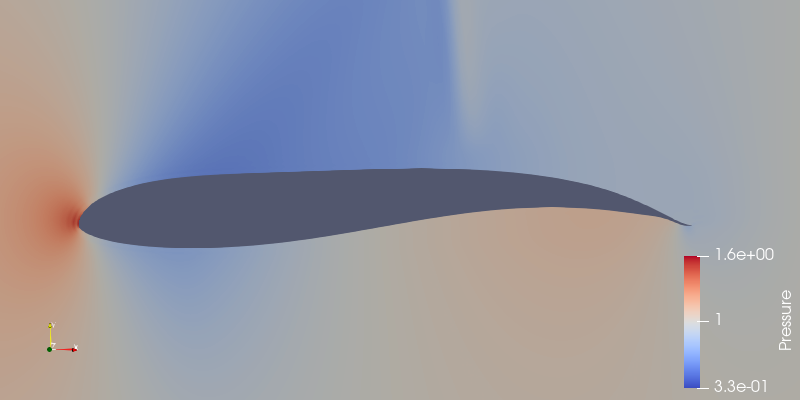}
    \end{subfigure}\hfill
    \begin{subfigure}[b]{\mysubfigwidth}
      \includegraphics[width=\linewidth]{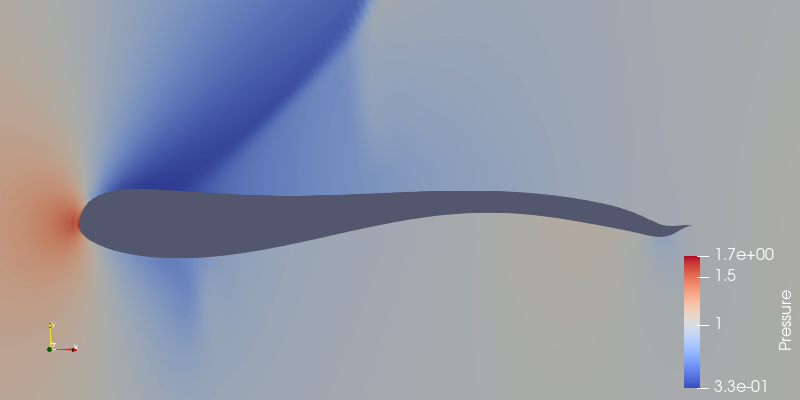}
    \end{subfigure}
    \subcaption{Bounds $(\underline{C_l}, \underline{t_c}) = (0.50,\,0.120)$}
  \end{subfigure}

  \caption{SLSQP optimized shape and flow results (pressure) for different lower constraint bounds $(\underline{C_l}, \underline{t_c}) \in \{0.3, 0.4, 0.5\} \times \{0.12\}$, comparing DM vs.\ HHM (left/right). Initialized with NACA0012.}
  \label{fig:slsqp_naca_t2}
\end{figure}

\subsection{Results Compared to Scaled Hicks-Henne Parameteratization with Physics Insight}
\label{sec:slsqp_constrained_hhm}

In this set of experiments, we compared our method with scaled HHM parameterizations resulting from engineering experiences and physics insights. In particular, we fix the first and last 5\% of the chord line—corresponding to the leading and trailing edges of the airfoil—throughout the shape optimization process. This common practice serves multiple purposes. From a geometric standpoint, these regions are highly sensitive and prone to numerical instabilities; small changes can introduce curvature discontinuities or unrealistic features that degrade mesh quality. Moreover, the leading edge plays a critical role in establishing proper stagnation behavior, while the trailing edge is essential for determining wake characteristics. Allowing unrestricted deformation in these regions may lead to ill-posed flow problems or solver divergence. Structurally, the leading edge is often reinforced and the trailing edge typically contains control surfaces or is limited by fabrication constraints due to its thin geometry. By keeping the first and last 5\% of the chord fixed, we ensure smoother deformations, preserve mesh quality, and maintain physically meaningful flow behavior, thereby enhancing both the robustness and reliability of the optimization process. Since certain parameters were fixed, this approach is considered a scaled Hicks-Henne parameterization method.

As shown in Figures \ref{fig:scaled_hhm2}, the up-tilting effects at the trailing edges were fixed with the introduction of variable scaling. The design performance of the scaled HHM approach improves substantially upon its baseline (comparing Table \ref{tab:scaled_HHM} with HHM results in Tables \ref{tab:rae_slsqp} and \ref{tab:naca_slsqp}).  Even though, the DM-based approach consistently outperforms the scaled HHM approach, as evidenced by comparing the DM results in Table \ref{tab:rae_slsqp} and \ref{tab:naca_slsqp} against Table \ref{tab:scaled_HHM}. The reductions in drag counts range from $1.3 - 34.2$ for the initialization with RAE2822 and $ 8.1 - 258.8 $ for the initialization with NACA0012. We note, scaled HHM method still requires the tuning of scalings for function and gradients. Some qualitative observations can be made:
\begin{itemize}
    \item As the optimization problem gets more challenging, i.e., with increasingly restrictive constraints, the performance advantage of the DM-based method increases. 
    \item The performance advantage of DM is more pronounced when initialized with NACA0012 versus RAE2822.
    \item Even for the simplest optimization problem, DM results in better performing shapes. 
\end{itemize}
These observations confirm the robustness and efficiency of our proposed manifold constrained optimization framework with a diffusion model.

\begin{table}[H]
  \centering
  \caption{Results of scaled HHM with RAE2822 and NACA0012 initializations and SLSQP Optimizer}
  \label{tab:scaled_HHM}
  \begin{tabular}{|cc|cc|cc|cc|cc|cc|}
    \hline
    \multicolumn{2}{|c|}{Problem}
      & \multicolumn{2}{c|}{\# $J$}
      & \multicolumn{2}{c|}{\# $\nabla J$}
      & \multicolumn{2}{c|}{$C_l$}
      & \multicolumn{2}{c|}{$t_c$}
      & \multicolumn{2}{c|}{$C_d$ (count)} \\
    \cline{1-2} \cline{3-4} \cline{5-6} \cline{7-8} \cline{9-10} \cline{11-12}
      $\underline{C_l}$ & $\underline{t_c}$
      & RAE & NACA
      & RAE & NACA
      & RAE & NACA
      & RAE & NACA
      & RAE & NACA  \\
    \hline
    0.30 & 0.105 & 14 & 59 & 14 & 27 & 0.3000 & 0.3000 & 0.1050 & 0.105 & 222.0 & 222.7 \\
    0.30 & 0.120 & 18 & 37 & 18 & 25 & 0.2993 & 0.3000 & 0.1197 & 0.1200 & 319.3 & 319.9 \\
    0.40 & 0.105 & 19 & 19 & 18 & 19 & 0.3997 & 0.4000 & 0.1049 & 0.1050 & 288.0 & 356.8 \\
    0.40 & 0.120 & 122 & 74 & 31 & 53 & 0.3996 & 0.5000 & 0.1198 & 0.1201 &  428.2 & 422.7 \\
    0.50 & 0.105 & 19 & 31 & 18 & 23 & 0.4985$^*$& 0.5000 & 0.1047$^*$& 0.1049 & 426.5 & 521.4 \\
    0.50 & 0.120 & 70 & 27 & 46 & 24 & 0.4998 & 0.5000 & 0.1200 & 0.1200 & 582.9 & 835.5 \\
    \hline
  \end{tabular}
\end{table}





\setlength{\mysubfigwidth}{0.45\textwidth}

\begin{figure}[htp]
  \centering

  \begin{subfigure}[b]{\textwidth}
    \centering
    \begin{subfigure}[b]{\mysubfigwidth}
      \includegraphics[width=\linewidth]{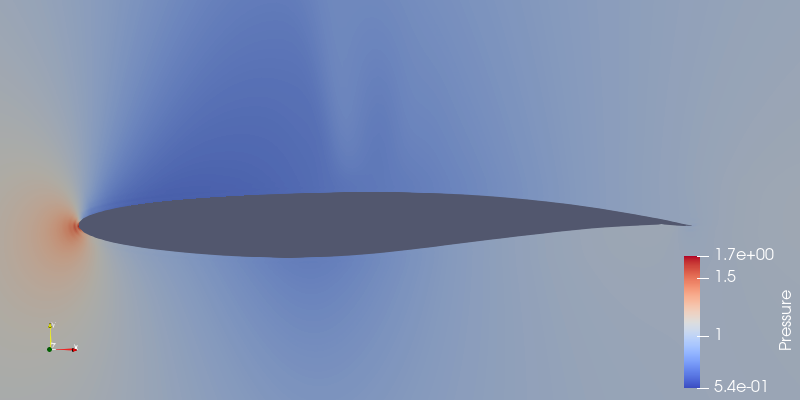}
      \subcaption{Bounds $(\underline{C_l}, \underline{t_c}) = (0.30,\,0.105)$}
    \end{subfigure}\hfill
    \begin{subfigure}[b]{\mysubfigwidth}
      \includegraphics[width=\linewidth]{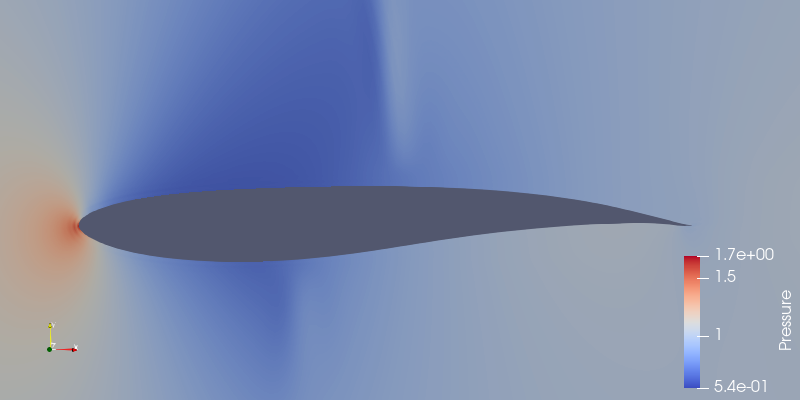}
      \subcaption{Bounds $(\underline{C_l}, \underline{t_c}) = (0.30,\,0.120)$}
    \end{subfigure}
  \end{subfigure}
  \medskip

  \begin{subfigure}[b]{\textwidth}
    \centering
    \begin{subfigure}[b]{\mysubfigwidth}
      \includegraphics[width=\linewidth]{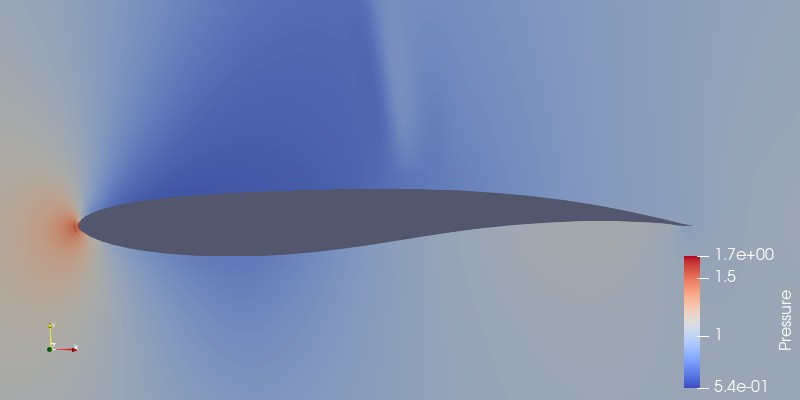}
      \subcaption{Bounds $(\underline{C_l}, \underline{t_c}) = (0.40,\,0.105)$}
    \end{subfigure}\hfill
    \begin{subfigure}[b]{\mysubfigwidth}
      \includegraphics[width=\linewidth]{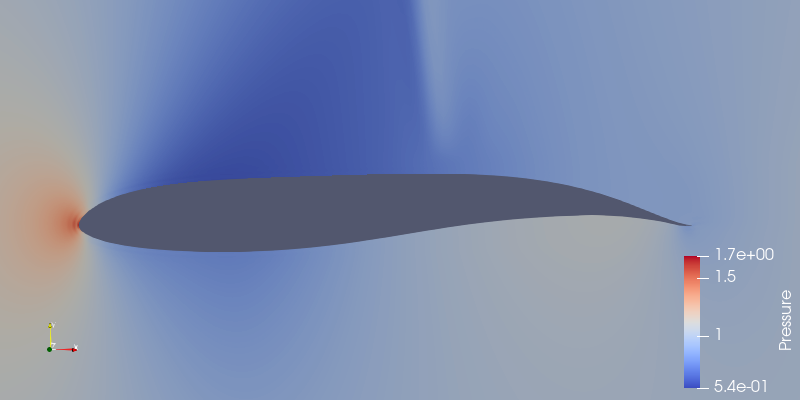}
      \subcaption{Bounds $(\underline{C_l}, \underline{t_c}) = (0.40,\,0.120)$}
    \end{subfigure}
  \end{subfigure}
  \medskip

  \begin{subfigure}[b]{\textwidth}
    \centering
    \begin{subfigure}[b]{\mysubfigwidth}
      \includegraphics[width=\linewidth]{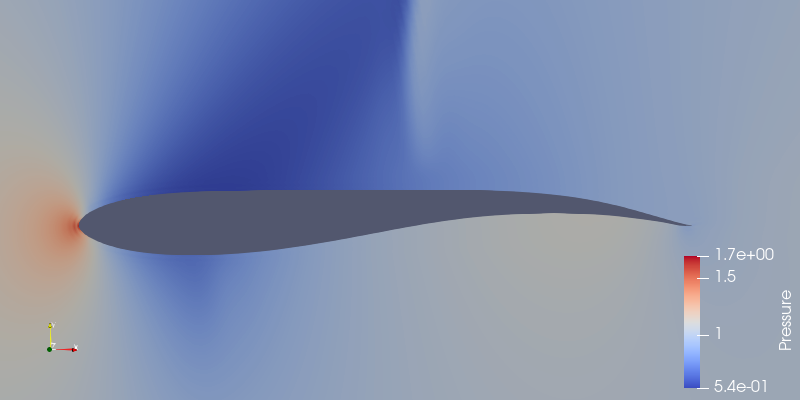}
      \subcaption{Bounds $(\underline{C_l}, \underline{t_c}) = (0.50,\,0.105)$}
    \end{subfigure}\hfill
    \begin{subfigure}[b]{\mysubfigwidth}
      \includegraphics[width=\linewidth]{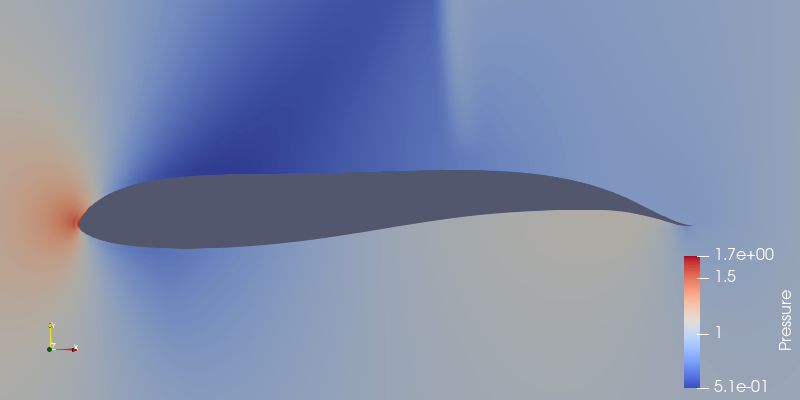}
      \subcaption{Bounds $(\underline{C_l}, \underline{t_c}) = (0.50,\,0.120)$}
    \end{subfigure}
  \end{subfigure}

  \caption{SLSQP optimized shape and flow results (pressure) for different lower constraint bounds $(\underline{C_l}, \underline{t_c}) \in \{0.3, 0.4, 0.5\} \times \{0.105, 0.120\}$. Initialized with RAE2822 and with scaled HHM.}
  \label{fig:scaled_hhm1}
\end{figure}


\setlength{\mysubfigwidth}{0.45\textwidth}

\begin{figure}[htp]
  \centering

  \begin{subfigure}[b]{\textwidth}
    \centering
    \begin{subfigure}[b]{\mysubfigwidth}
      \includegraphics[width=\linewidth]{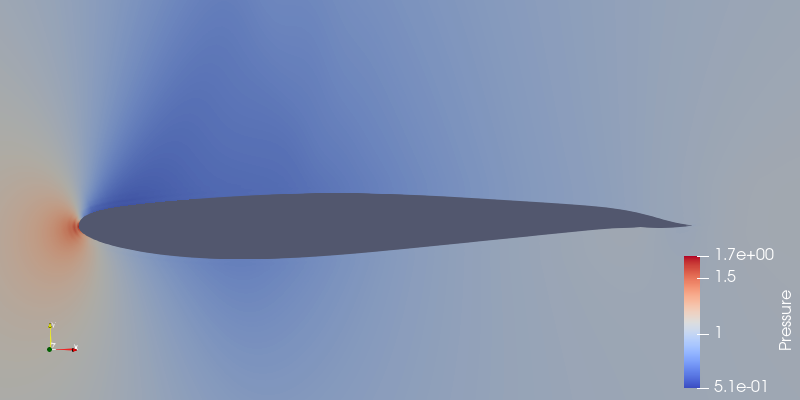}
      \subcaption{Bounds $(\underline{C_l}, \underline{t_c}) = (0.30,\,0.105)$}
    \end{subfigure}\hfill
    \begin{subfigure}[b]{\mysubfigwidth}
      \includegraphics[width=\linewidth]{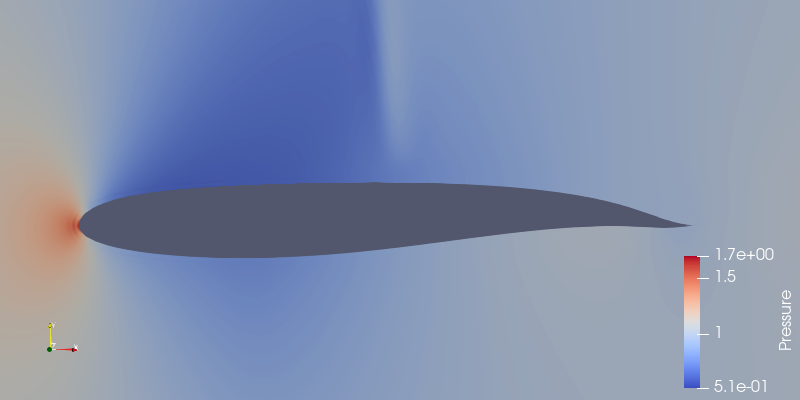}
      \subcaption{Bounds $(\underline{C_l}, \underline{t_c}) = (0.30,\,0.120)$}
    \end{subfigure}
  \end{subfigure}
  \medskip

  \begin{subfigure}[b]{\textwidth}
    \centering
    \begin{subfigure}[b]{\mysubfigwidth}
      \includegraphics[width=\linewidth]{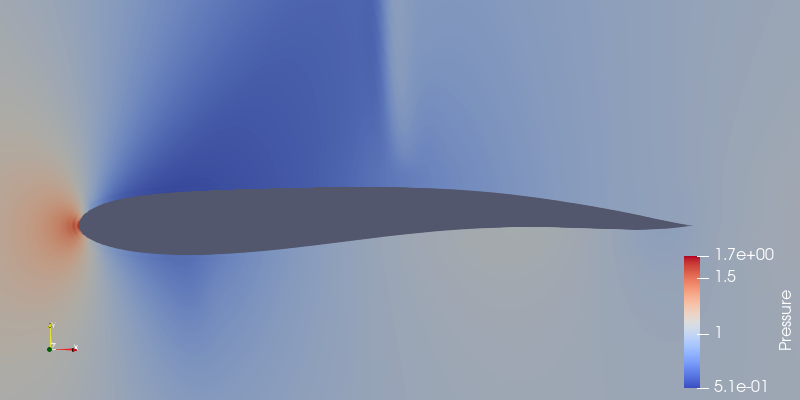}
      \subcaption{Bounds $(\underline{C_l}, \underline{t_c}) = (0.40,\,0.105)$}
    \end{subfigure}\hfill
    \begin{subfigure}[b]{\mysubfigwidth}
      \includegraphics[width=\linewidth]{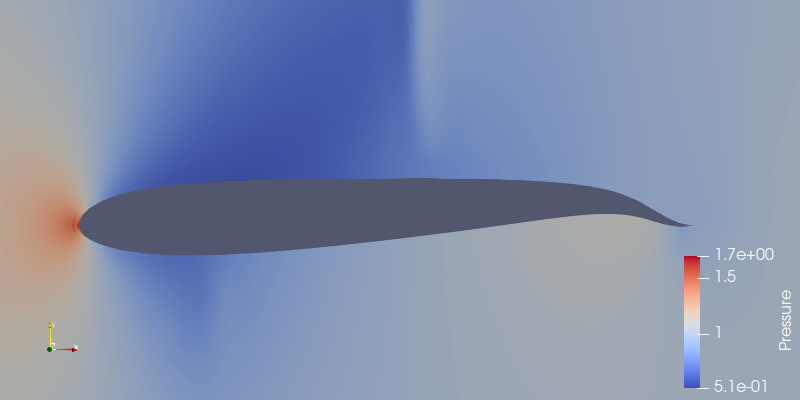} 
      \subcaption{Bounds $(\underline{C_l}, \underline{t_c}) = (0.40,\,0.120)$}
    \end{subfigure}
  \end{subfigure}
  \medskip

  \begin{subfigure}[b]{\textwidth}
    \centering
    \begin{subfigure}[b]{\mysubfigwidth}
      \includegraphics[width=\linewidth]{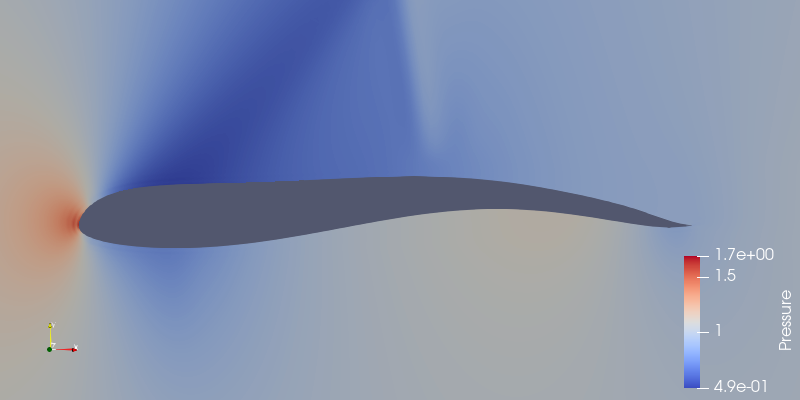}
      \subcaption{Bounds $(\underline{C_l}, \underline{t_c}) = (0.50,\,0.105)$}
    \end{subfigure}\hfill
    \begin{subfigure}[b]{\mysubfigwidth}
      \includegraphics[width=\linewidth]{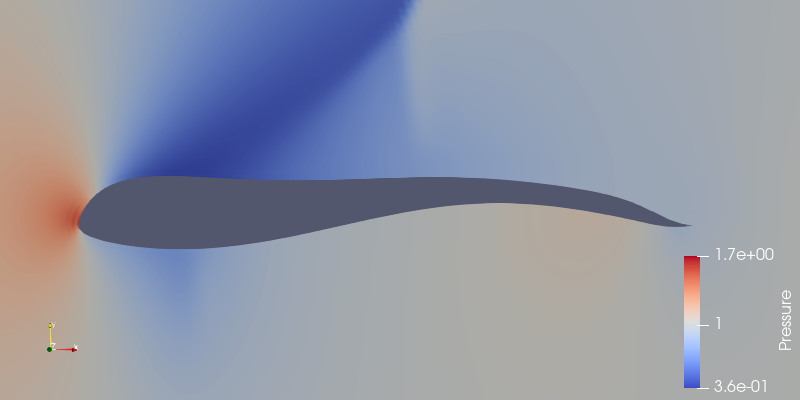}
      \subcaption{Bounds $(\underline{C_l}, \underline{t_c}) = (0.50,\,0.120)$}
    \end{subfigure}
  \end{subfigure}

  \caption{SLSQP optimized shape and flow results (pressure) for different lower constraint bounds $(\underline{C_l}, \underline{t_c}) \in \{0.3, 0.4, 0.5\} \times \{0.105, 0.120\}$. Initialized with NACA0012 and with scaled HHM.}
  \label{fig:scaled_hhm2}
\end{figure}


\clearpage

\subsection{Results with NACA0012 Initialization and IPOPT Optimizer}
\label{sec:ipopt_naca}

In this set of experiments, we assess the robustness of our proposed method with two very different gradient-based optimizers: SLSQP (a sequential quadratic programming solver) and IPOPT (an interior-point method solver). By default, IPOPT evaluates both the objective and its gradient at every new design, while SLSQP evaluates functions more often than gradients due to its line search implementation. As shown in Table \ref{tab:solvers_naca}, except for the test case $(\underline{C_l}, \underline{t_c})=(0.30, 0.120)$, SLSQP finds slightly better solutions. For the problem $(0.40, 0.120)$, IPOPT seems to be trapped in an inferior local solution with a siginifantly higher drag counts. 

Importantly, our proposed method works ``out of box'' with both optimizers, without requiring manual tuning of solver or optimization problem parameters. In contrast, with HHM we were unable to find a working parameter set for IPOPT that gives useful optimized shapes. Figure \ref{fig:ipopt_naca} shows the airfoil shapes obtained with IPOPT and DM, which differ noticeable from those obtained using SLSQP and DM in Figures \ref{fig:slsqp_naca_t1} and \ref{fig:slsqp_naca_t2}, although their thickness, lift, and drag are nearly identical. This variety of equally good-performing shapes provides evidence of the benign optimization landscape on the learned manifold. Furthermore, multiple designs grant engineers extra flexibility.

\begin{table}[H]
  \centering
  \caption{Comparison of optimizers SLSQP and IPOPT with DM method with NACA0012 initialization}
  \label{tab:solvers_naca}
  \begin{tabular}{|cc|cc|cc|cc|cc|cc|}
    \hline
    \multicolumn{2}{|c|}{Problem}
      & \multicolumn{2}{c|}{\# $J$}
      & \multicolumn{2}{c|}{\# $\nabla J$}
      & \multicolumn{2}{c|}{$C_l$}
      & \multicolumn{2}{c|}{$t_c$}
      & \multicolumn{2}{c|}{$C_d$ (count)} \\
    \cline{1-2} \cline{3-4} \cline{5-6} \cline{7-8} \cline{9-10} \cline{11-12}
      $\underline{C_l}$ & $\underline{t_c}$
      & SLSQP & IPOPT
      & SLSQP & IPOPT
      & SLSQP & IPOPT
      & SLSQP & IPOPT
      & SLSQP & IPOPT
      \\
    \hline
    0.30 & 0.105 & 37 &   20   & 33 &  20    & 0.3001 &   0.3010    & 0.1050 &   0.1055    & 214.6 &   230.5     \\
    0.30 & 0.120 & 12 &  17  & 11 &   17   & 0.3000 &   0.3000    & 0.1199 &    0.1200   & 311.7 &    311.0    \\
    0.40 & 0.105 & 10 &   14  &  9 &  14    & 0.4000 &   0.3997    & 0.1049 &    0.1050   & 296.5 &    290.5    \\
    0.40 & 0.120 & 24 &   5   & 16 &   5   & 0.4002 &  0.3966   & 0.1199 & 0.1199  & 402.9 &   549.0     \\
    0.50 & 0.105 & 10 &   12   &  8 &   12   & 0.4997 &   0.4988    & 0.1049 &   0.1049    & 388.9 &   390.1     \\
    0.50 & 0.120 & 62 &   30   & 36 &   30   & 0.4998 &  0.4987     & 0.1198 &   0.1205    & 576.7 &   582.3     \\
    \hline
  \end{tabular}
\end{table}


\setlength{\mysubfigwidth}{0.45\textwidth}

\begin{figure}[htp]
  \centering

  \begin{subfigure}[b]{\textwidth}
    \centering
    \begin{subfigure}[b]{\mysubfigwidth}
      \includegraphics[width=\linewidth]{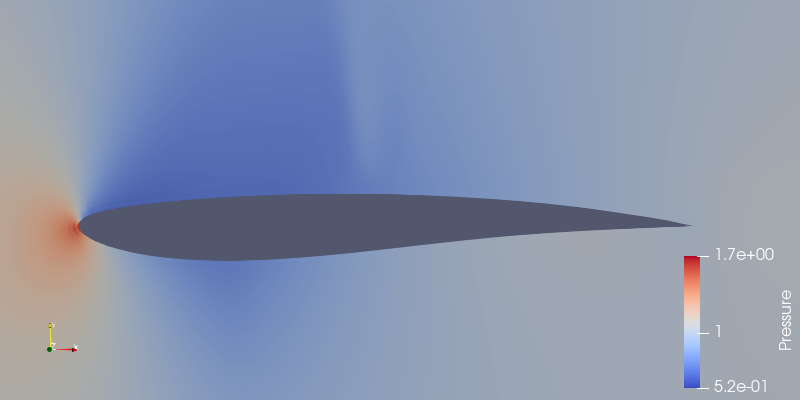}
      \subcaption{Bounds $(\underline{C_l}, \underline{t_c}) = (0.30,\,0.105)$}
    \end{subfigure}\hfill
    \begin{subfigure}[b]{\mysubfigwidth}
      \includegraphics[width=\linewidth]{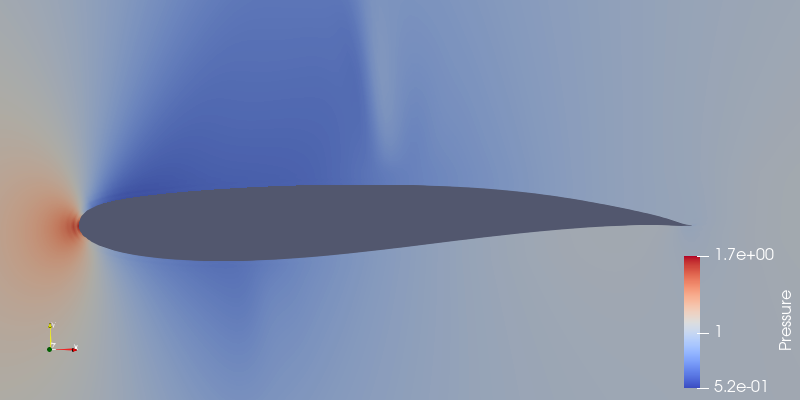}
      \subcaption{Bounds $(\underline{C_l}, \underline{t_c}) = (0.30,\,0.120)$}
    \end{subfigure}
  \end{subfigure}
  \medskip

  \begin{subfigure}[b]{\textwidth}
    \centering
    \begin{subfigure}[b]{\mysubfigwidth}
      \includegraphics[width=\linewidth]{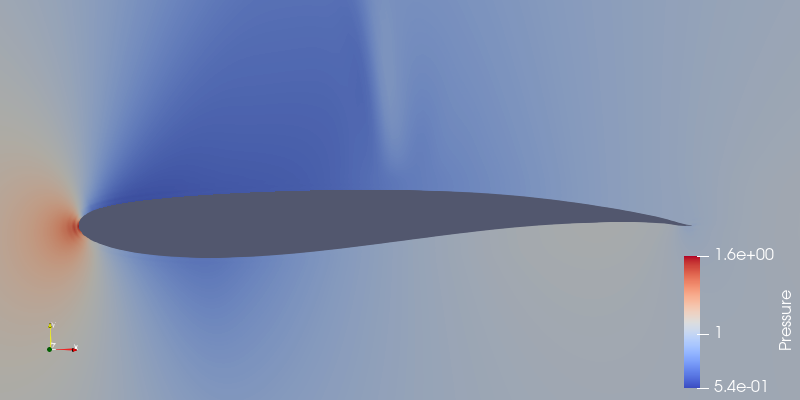}
      \subcaption{Bounds $(\underline{C_l}, \underline{t_c}) = (0.40,\,0.105)$}
    \end{subfigure}\hfill
    \begin{subfigure}[b]{\mysubfigwidth}
      \includegraphics[width=\linewidth]{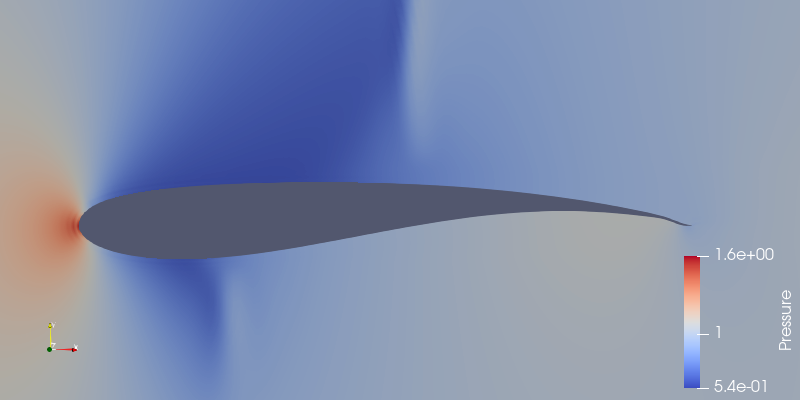}
      \subcaption{Bounds $(\underline{C_l}, \underline{t_c}) = (0.40,\,0.120)$}
    \end{subfigure}
  \end{subfigure}
  \medskip

  \begin{subfigure}[b]{\textwidth}
    \centering
    \begin{subfigure}[b]{\mysubfigwidth}
      \includegraphics[width=\linewidth]{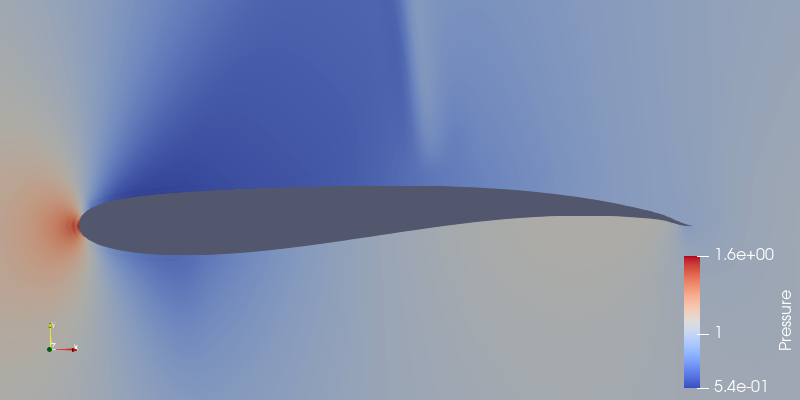}
      \subcaption{Bounds $(\underline{C_l}, \underline{t_c}) = (0.50,\,0.105)$}
    \end{subfigure}\hfill
    \begin{subfigure}[b]{\mysubfigwidth}
      \includegraphics[width=\linewidth]{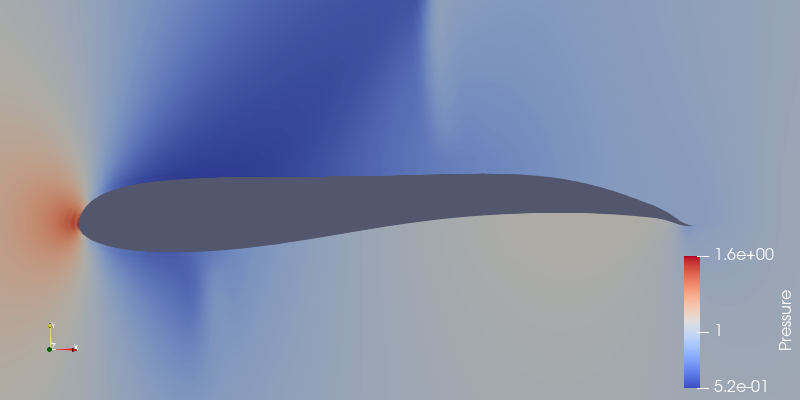}
      \subcaption{Bounds $(\underline{C_l}, \underline{t_c}) = (0.50,\,0.120)$}
    \end{subfigure}
  \end{subfigure}

  \caption{IPOPT optimized shape and flow results (pressure) for different lower constraint bounds $(\underline{C_l}, \underline{t_c}) \in \{0.3, 0.4, 0.5\} \times \{0.105, 0.120\}$. Initialized with NACA0012.}
  \label{fig:ipopt_naca}
\end{figure}

\section{Conclusions}

Significant advances in adjoint methods have enabled efficient and accurate computation of shape gradients for aerodynamic shape optimization problems. However, solving these optimization problems remains challenging, often requiring ad hoc tuning of optimization parameters and variable scaling. To address this, we propose a framework that constrains the design space to a manifold of aerodynamic shapes. Since no explicit, first-principle based mathematical formulation exists for defining such a manifold, we implicitly learn it using a diffusion model trained with existing high-performing shapes. We derive and implement a fully differentiable framework that backpropagates shape adjoints onto the latent space of this manifold, enabling the application of gradient-based optimization algorithms. Extensive computational experiments demonstrate that our proposed approach eliminates ad hoc parameter tuning and variable scaling, maintains robustness across initialization and optimization solver choices, and achieves superior aerodynamic performance compared to conventional approaches. These results confirm that the learned manifold constraint significantly improves the gradient-based optimization search. Moreover, the approach introduces minimal computational overhead, and can be easily integrated into existing adjoint-based aerodynamic shape optimization workflows. More broadly, incorporating diffusion models into optimization has become a growing trend in recent research \cite{li2024diffusion}. This work establishes how diffusion models can synergize effectively with adjoint-based aerodynamic shape optimization and motivates further exploration of their application to challenging engineering design problems.
    
\color{black}

\bibliography{sample}

\end{document}